\documentclass[natbib,sn-basic]{sn-jnl}

\jyear{2023}

\begin{document}

\title[Observationally guided models for the solar
  dynamo]{Observationally guided models for the solar
  dynamo and the role of the surface field}

\author{\fnm{Robert H.} \sur{Cameron}}

\author{\fnm{Manfred} \sur{Sch{\"u}ssler}}

\affil{\orgname{Max Planck Institute for Solar System Research},
\orgaddress{\street{Justus-von-Liebig-Weg 3}, \city{37077 G{\"o}ttingen}, 
\country{Germany}}}

\abstract{Theoretical models for the solar dynamo range from simple
  low-dimensional ``toy models'' to complex 3D-MHD simulations. Here
  we mainly discuss appproaches that are motivated and guided by solar
  (and stellar) observations. We give a brief overview of the
  evolution of solar dynamo models since 1950s, focussing upon the
  development of the Babcock-Leighton approach between its
  introduction in the 1960s and its revival in the 1990s after being
  long overshadowed by mean-field turbulent dynamo theory. We
  summarize observations and simple theoretical deliberations that
  demonstrate the crucial role of the surface fields in the dynamo
  process and and give quantitative analyses of the generation and
  loss of toroidal flux in the convection zone as well as of the
  production of poloidal field resulting from flux emergence at the
  surface. Furthermore, we discuss possible nonlinearities in the
  dynamo process suggested by observational results and present
  models for the long-term variability of solar activity motivated by
  observations of magnetically active stars and the inherent
  randomness of the dynamo process.}

\keywords{solar activity, solar cycle, dynamo}

\maketitle

\section{Introduction}\label{sec:intro:MS}

Studies of solar and stellar dynamos face a problem of utter
complexity, i.e., the interaction of turbulent convection with
rotation and magnetic field in a highly stratified medium, covering
wide ranges of spatial and temporal scales. Attempts to directly
attack this problem with numerical 3D-MHD simulations have made
significant progress in recent years \citep[e.g.,][]{Charbonneau:2020,
  Browning:etal:2023}, but are still severely limited in spatial
resolution, so that a faithful representation of the solar cycle has
not been achieved so far. All other approaches to the dynamo problem
resort to simplifications in order to obtain a tractable task. They
largely rely on the surprising amount of regularity shown by the solar
cycle, e.g., Hale's polarity rules and the systematic tilt of sunspot
groups (Joy's law), the butterfly diagram, and the regular reversals
of the global dipole field \citep{Hathaway:2015}. These regularities
suggest simplified concepts, such as the generation of azimuthal
(toroidal) magnetic flux through winding of meridional (poloidal)
magnetic field lines by differential rotation \citep{Cowling:1953} or
the formation of bipolar sunspot groups by the emergence of buoyantly
rising magnetic flux tubes \citep{Parker:1955a}.

A variety of models addressing various aspects of the dynamo problem
has been developed. These range from rather ad-hoc ``toy models'' over
sophisticated two-scale approaches, pioneered by the model of
\citet{Parker:1955b} and by mean-field theory of turbulent MHD flows
\citep[see review by][]{Brandenburg:etal:2023} to 2D/3D
flux-transport dynamo models including various physical processes
considered to be relevant for the dynamo \citep[see reviews
  by][]{Charbonneau:2020, Hazra:etal:2023}.
 
The simplest ``model'' in the literature is probably due to
\citet{Barnes:etal:1980}. These authors considered a digital
narrow-band filtering of Gaussian white noise, corresponding to a
randomly disturbed periodic signal that could be illustrated by a
``pendulumn pelted with peas'' \citep{Yule:1927}. The results of their
short computer program (15 lines!) exhibit a very similar kind of
variability as shown by the sunspot record, including the occurence of
extended periods of low activity (grand minima). The important lesson
from this result for models of the solar dynamo is that even a
striking similarity of their output with records of solar activity alone
does not necessarily imply the validity of a model.

In this paper, we focus upon models of the solar dynamo as well as for
the cyclic generation and removal of magnetic flux that are guided by
solar (and also stellar) observations.  The motivation for such
approaches comes from the fact that the very complex interactions in
the convection zone can lead to comparably simple results, such as the
stable differential rotation in latitude, the meridional circulation,
the polarity rules of sunspot groups, or the butterfly diagram of
magnetic flux emergence.  This might perhaps be compared to the flow
of water in a watermill: although the detailed turbulent flow pattern
is utterly complex, the general result is simple: water flows down the
potential well and faithfully drives the wheel.  The paradigm of
observationally motivated models is the scenario of
\citet{Babcock:1961}, which was later extended and put in a
mathematical form by \citet{Leighton:1969}. These seminal papers paved
the way for what are now generally called Babcock-Leighton-type models
of the solar dynamo.

The plan of this paper is as follows. In Sec.~\ref{sec:BL} we describe
the principles of the Babock-Leighton scenario and review the
evolution of dynamo models until the end of the 1980s, when mounting
empirical evidence led to the renaissance of Babcock-Leighton dynamo
models after an extended period of near oblivion. The generation and
removal of toroidal and poloidal magnetic flux in the Sun and the
crucial role of the observable surface field are discussed in
Sec.~\ref{sec:gen}. Nonlinearity of the dynamo process, predictability, and models for the long-term variability of solar activity are considered in
Sec.~\ref{sec:fluct}. Sec.~\ref{sec:outlook} gives a brief outlook.

\section{The Babcock-Leighton approach and the evolution of models for 
         the solar dynamo}
\label{sec:BL}

The seminal paper of \citet{Babcock:1961} lays out a purely
observationally based scenario for the 11/22-year solar cycle. In a
first step, the poloidal magnetic flux connected to the global dipole
field is wound up by latitudinal differential rotation in the
convection zone. This generates oppositely directed subsurface belts
of toroidal field in both hemispheres.  Subsequent instability,
buoyant rise, and emergence of toroidal flux at the surface produces
bipolar magnetic regions (BMRs) and sunspot groups in accordance with
Hale's polarity rules. These regions are observed to be systematically
tilted in the sense that the magnetic polarity leading in the
direction of rotation is nearer to the equator than the following
polarity. This tilt, together with the dispersal of the BMRs in the
course of time, leads to preferred transport of leading-polarity flux
over the equator, so that a surplus of the respective
following-polarity flux builds up in both hemispheres. Poleward
transport of this flux by convection and meridional circulation leads
to the reversal of the polar fields and the buildup of an oppositely
directed dipole field, a process already suggested by
\citet{Babcock:Babcock:1955}. This entails the generation and
emergence of a reversed toroidal field followed by another reversal of
the dipole field, thus giving rise to the 22-year magnetic cycle.

\citet{Babcock:1961} already noted that the systematic tilt of the
sunspot groups could be ``... the result of Coriolis forces, which
induce a vorticity in the whole of the fluid associated with a BMR
when it rises to the surface.'' The Coriolis effect is also invoked in
the concept of ``cyclonic convection'' of \citet{Parker:1955b} and in
the mean-field scheme based on turbulence theory pioneered by
\citet{Steenbeck:Krause:1966} and \citet{Steenbeck:etal:1966}.
Possibly since both these approaches focus on the collective effect of
a small-scale process (in the sense of a two-scale approach), neither
Parker nor Steenbeck and coworkers apparently realized that the systematic
tilt provides direct observational support for the action of the
Coriolis effect as well a quantitative measure.

\citet{Leighton:1964} added a key component to Babcock's scenario in
the form of a random-walk model for the quasi-diffusive transport of
magnetic flux on the solar surface by supergranular
convection. Subsequently, he put the Babcock model in the mathematical
form of a one-dimensional dynamo equation \citep{Leighton:1969}. It is
not obvious why this Babcock-Leighton (BL) model fell into near
oblivion for about two decades thereafter. In order to provide some
understanding for this development, a brief sketch of the evolution
of solar dynamo studies from the early 1970s onward is given in what
follows.

Perhaps since the mathematically involved mean-field theory was
considered to provide a strong theoretical footing, models for the
solar dynamo until the 1990s mostly relied on a mean-field turbulent
``$\alpha$-effect'' based on correlations between small-scale fields
operating within the convection zone. To explain the
equatorward-directed migration of the activity belts shown by the
butterfly diagram, these models generally required an inwardly
increasing rotation rate, $d\Omega/dr<0$ \citep{Koehler:1973,
  Yoshimura:1975}. The effect of the observed strong latitudinal
differential rotation was thought to be of secondary importance.
Consequently, both key processes for the turbulent dynamo, radial
differential rotation and $\alpha$-effect, were assumed to operate in
the solar interior, thus inaccessable to observation at that
time. Therefore, the corresponding model parameters were largely
unconstrained. The observable evolution of the surface field was seen
as a mere epiphenomenon of a dynamo process operating deeply hidden in
the convection zone and served solely as an observational constraint
for the models.

The foundations of this approach were undermined when helioseismology
made it possible to measure the differential rotation in the
convection zone. It turned out that (except for a shallow near-surface
shear layer) the increase of the rotation rate with depth required by
these models is absent in the bulk of the convection zone \citep[see
  review by][]{Howe:2009}. On the other hand, the rotation rate at the
bottom of the convection zone showed a steep radial gradient across
the ``tachocline'' \citep{Brown:etal:1989}, such that $d\Omega/dr<0$
in solar latitudes $\ge 45\,$deg and $d\Omega/dr>0$ in lower
latitudes. Some years before, \citet{Galloway:Weiss:1981} already had
suggested that the solar dynamo should work in a stably stratified
layer of overshooting convection below the bottom of the convection
zone in order to avoid the rapid buoyant loss of toroidal magnetic
flux inferred by \citet{Parker:1975}. These results led to the concept
of turbulent dynamo action within the overshoot layer/tachocline
\citep[e.g.,][]{Ruediger:Brandenburg:1995} or near its interface with
the convection zone proper \citep{Parker:1993}, where a properly
chosen combination of the signs of $\alpha$-effect and radial gradient
of the rotation rate could provide the correct conditions for
equatorward propagating dynamo waves and activity belts
\citep[e.g.,][]{Tobias:1996,Charbonneau:MacGregor:1997}.  This
approach was complemented by studies of equilibrium, stability, and
dynamics of thin magnetic flux tubes starting with
\citet{Spruit:Ballegooijen:1982}, \citet{Choudhuri:Gilman:1987}, and
\citet{Moreno-Insertis:etal:1992}. Such studies \citep[see reviews
  by][]{Fan:2021,Isik:etal:2023} indicated that the toroidal field in
the overshoot region must be amplified to about $10^5\,$G before
becoming unstable \citep{Ferriz-Mas:Schuessler:1993,
  Ferriz-Mas:Schuessler:1995} and rising to the surface to form
bipolar magnetic regions within the latitude range shown by the
butterfly diagram \citep{Schuessler:etal:1994}.  Twisting of the
rising flux tubes due to the Coriolis force then leads to tilt angles
consistent with observation \citep{Fan:etal:1994, Caligari:etal:1995}.

Although the combination of tachocline/overshoot layer dynamos and the
dynamics of thin flux tubes offered a comprehensive picture from the
generation of magnetic flux up to its emergence at the surface, these
models relied on a number of untested assumptions and simplifications,
so that their predictive power remained rather limited.  In the course
of time, a number of theoretical considerations and observational
results cast severe doubt upon the validity of this aproach:

\begin{itemize}
\item{The total energy energy of a toroidal field of $10^5\,$Gauss at
  the bottom of the convection zone would be comparable to the energy
  in the differential rotation of the tachocline
  \citep{Rempel:2006}. However, a corresponding strong variation of
  the differential rotation in the lower convection zone and
  tachocline in the course of the 11-year activity cycle is not
  observed \citep{Basu:Antia:2019}.  Moreover, the interface between
  radiative core and convection zone cannot support much shear
  stress \citep{Spruit:2011}. Consequently, the radial differential
  rotation in the tachocline (mainly reflecting the transition from
  latitudinally differential rotation in the convection zone to nearly
  rigid rotation below) cannot generate a sizeable amount of toroidal
  magnetic flux unless being maintained by a very powerful downward
  transport of angular momentum.}
\item{The latitudinal differential rotation is sufficient to create
  the total toroidal flux covered by the emergence of bipolar magnetic
  regions and sunspot groups. At the same time, the contribution of
  the observed radial differential rotation in the convection zone is
  only a few percent of that of the latitudinal differential rotation
  \citep{Cameron:Schuessler:2015}. Moreover, the radial shear in the
  tachocline is strong in high latitudes and weak in low latitudes,
  thus unfavorable for toroidal flux generation in low latitudes. }
\item{Helioseismology indicates that the overshoot layer is much more
  strongly subadiabatic than assumed in the models for the storage and
  stability of toroidal flux \citep{Christensen-Dalsgaard:etal:2011}.
  On the other hand, part of the lower convection zone could be
  subadiabatically stratified \citep[e.g.,][]{Spruit:1997, Hotta:2017}.}
\item{Observations of magnetically active stars show that partly and
  fully convective stars follow the same activity-rotation law
  \citep{Wright:Drake:2016, Reiners:etal:2022}, thus questioning the
  relevance of convective overshoot and the existence of a tachocline
  for the dynamo process.  Furthermore, even very cool, fully
  convective dwarfs (beyond spectral type M7) can exhibit activity
  cycles \citep{Route:2016}.}
\item{3D-MHD simulations demonstrate the formation of
  super-equipartion magnetic flux concentrations and buoyantly rising
  flux loops within a simulated convection zone without overshoot and
  tachocline \citep{Nelson:etal:2014, Fan:Fang:2014, Chen:etal:2017}.}
\item{If emerged magnetic structures were ``anchored'' near to the
  bottom of the convection zone, they should show slower rotation 
  than that of the surface plasma below about $\pm 30\,$deg
  latitude according to the helioseismically determined rotation
  profile. However, magnetic structures are observed
  rotate faster than the plasma at the surface \citep[e.g.,][]{Howard:1996}.}
\end{itemize}

A new twist came when observations of a systematic poleward-directed
meridional flow at the surface pioneered by \citet{Duvall:1979} and
\citet{Howard:1979} became established during the 1980s \citep[see
  review by][]{Hanasoge:2022}.  This led to the suggestion that the
associated deep return flow within the convection zone could transport
toroidal flux equatorward.  Models of flux-transport dynamos (FTDs)
that rely upon this concept can provide a butterfly diagram consistent
with observations regardless of the sign of the radial gradient of the
rotation rate \citep{Wang:Sheeley:1991, Durney:1995,
  Choudhuri:etal:1995}. A recent review of FTD models has been provided by
\citet{Hazra:etal:2023}.

The confirmation of the systematic poleward surface flow also sparked
the development of simulation models for the transport of magnetic
flux at the solar surface on solar-cycle time scales, pioneered by the
NRL group \citep{DeVore:etal:1984,Wang:etal:1989a}. Such Surface Flux
Transport (SFT) models \citep[see reviews by][]{Mackay:Yeates:2012,
  Yeates:etal:2023} assume passive transport of vertically (radially)
orientated magnetic flux by surface flows. Comparison of SFT
simulations with observed synoptic magnetograms shows that the
evolution of the surface flux can be faithfully described by flux
emergence in systematically tilted bipolar magnetic regions followed
by passive flux transport by differential rotation, meridional flow,
and supergranular flows \citep[the latter mostly being treated as a
  diffusion-like random walk, cf.][]{Leighton:1964}, and flux
cancellation. In particular, the models confirm the buildup of polar
fields due to the preferred transport of following-polarity flux
toward the poles \citep{Wang:etal:1989b}.

The success of the SFT simulations led to the re-appraisal of the BL
approach for the (re)generation of the poloidal field in the course of
the dynamo process \citep{Giovanelli:1985, Wang:Sheeley:1991},
typically in connection with equatorward flux transport by meridional
flow in the convection zone as Flux Transport Dynamo (FTD) models
\citep{Wang:etal:1991, Charbonneau:2020, Hazra:etal:2023}. In the
course of time, further lines of evidence for the validity of the BL
approach and the crucial role of the surface flux for the solar dynamo
became apparent:
\begin{itemize}
\item{SFT models reproduce the evolution of the surface field,
  particularly the poleward drift of the following-polarity surface
  field leading to the buildup of the polar fields and the axial
  dipole \citep[e.g.,][]{Wang:etal:2002, Baumann:etal:2004,
    Jiang:etal:2014, Upton:Hathaway:2014, Whitbread:etal:2017}}.
\item{The polar field strength around cycle minimum is strongly
  correlated with the amplitude of the subsequent activity maximum
  \citep{Legrand:Simon:1981, Wang:Sheeley:2009,
    Kitchatinov:Olemskoy:2011, Hathaway:Upton:2016}, thus providing
  the most faithful predictor of cycle strength
  \citep{Schatten:etal:1978, Petrovay:2020, Kumar:etal:2021,
    Bhowmik:etal:2023}.  In addition to confirming this correlation,
  \citet{Munoz-Jaramillo:etal:2013} showed that the ``memory'' of the
  system does not extend beyond one cycle.}
\item{\citet{Cameron:Schuessler:2015} showed that the above
  correlation in fact reflects a causation: the net hemispheric
  toroidal flux generated during a half cycle results from the action
  of the latitudinal differential rotation on the poloidal flux
  connected to the polar fields (see Sec.~\ref{subsec:btor} below).}
\item{The observed azimuthal surface field (a proxy for flux
  emergence) evolves in accordance with an updated BL model
  \citep{CameronDuvall::2018}.}
\end{itemize}
The timescale for the rise of flux loops formed by the toroidal field
and the subsequent flux emergence is generally considered to be short
compared to the cycle time scale. Therefore, BL dynamo models
typically do not explicitely include these processes but rather
incorporate a source term near the surface that is related to the
deep-seated toroidal field in the convection zone \citep[for a
  discussion of various approaches, see][]{Choudhuri:Hazra:2016}.  A
few examples of two-dimensional axisymmetric dynamo models with
different prescriptions for the BL source term are the studies of
\citet{Durney:1997}, \citet{Dikpati:Charbonneau:1999},
\citet{Nandy:Choudhuri:2001}, \citet{Chatterjee:etal:2004}, and
\citet{Munoz:Jaramillo:etal:2010}. More complete overviews have been
provided by \citet{Charbonneau:2020} and \citet{Hazra:etal:2023}.

FTD/BL models mostly rely on radial differential rotation in the
tachocline and often also assume penetration of the meridional flow
into the stably stratified interior in order to ``store'' the toroidal
magnetic flux at the bottom of the convection zone
\citep[e.g.,][]{Nandy:Choudhuri:2002}.  In contrast, the 2D model of
\citet{Zhang:Jiang:2022} exhibits no such penetration, but
consistently includes the helioseismically determined differential
rotation in the convection zone (including the near-surface shear
layer), a one-cell meridional circulation, radial pumping keeping the
surface field vertical and inhibiting diffusive loss of the toroidal
field, and a BL source term.  The generation of toroidal flux turns
out to be strongly dominated by the latitudinal differential rotation
in the bulk of the convection zone \citep[see
  also][]{Guerrero:Gouveia:2007, Munoz-Jaramillo:etal:2009}. The
latitudinal propagation of the toroidal flux belts in the model of
\citet{Zhang:Jiang:2022} is provided by a combination of flux
transport by the equatorward meridional return flow and the latitude
dependence of the latitudinal rotational shear generating toroidal
magnetic flux, the latter as already envisaged by
\citet{Babcock:1961}. Test cases with removal of the radial shear in
the tachocline or putting the numerical boundary above the tachocline
does not significantly change the results of the model, thus strongly
suggesting that the tachocline shear is indeed largely irrelevant for
the dynamo \citep[cf.][]{Spruit:2011, Brandenburg:2005,
  Cameron:Schuessler:2015}.

\begin{figure}
\centering
\includegraphics[width=\hsize]{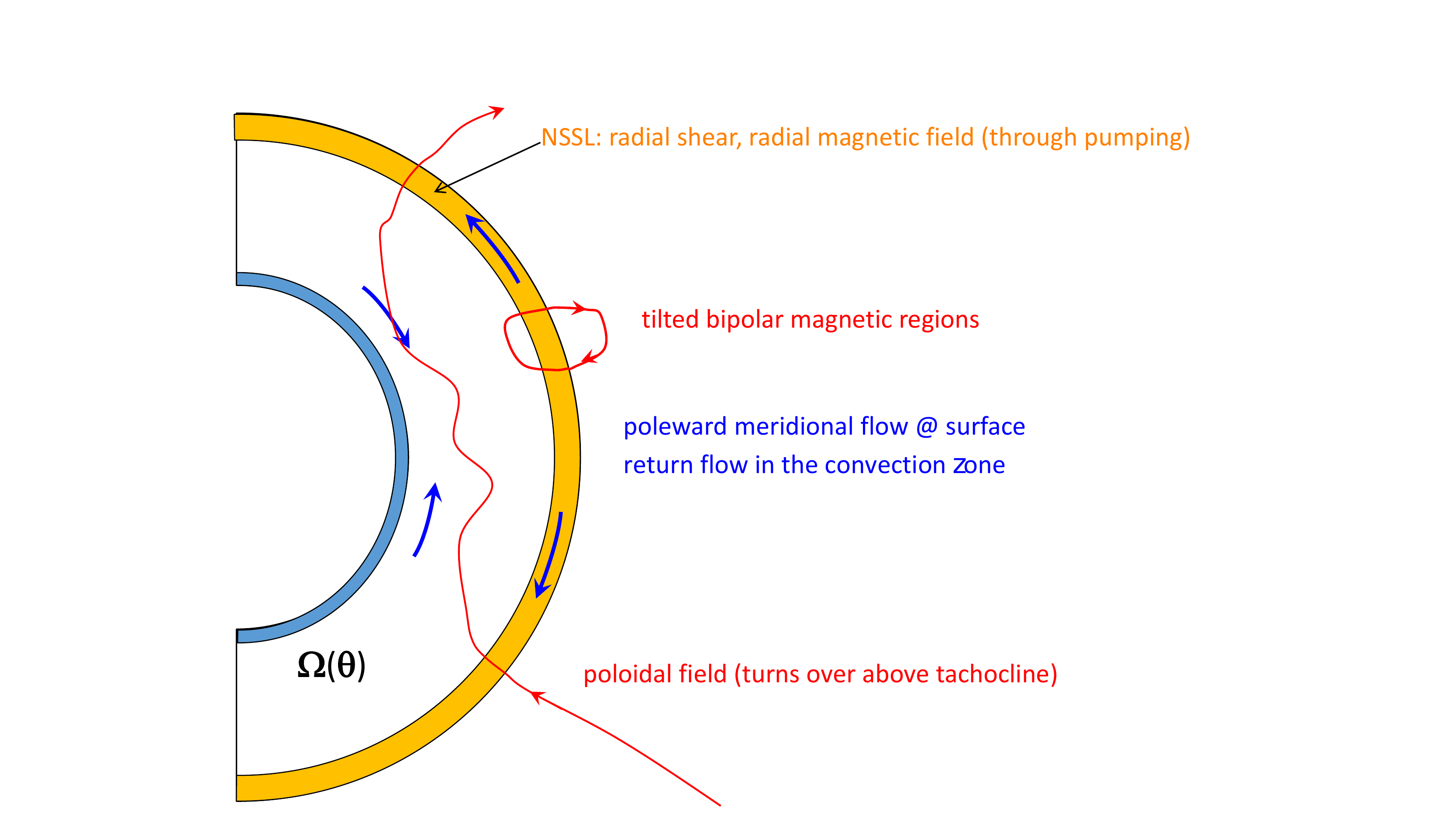}
\caption{Sketch of the updated Babcock-Leighton model of 
\citet{CameronSchussler::2017b}.}
\label{fig:CS2017a}
\end{figure}

\begin{figure}
\centering
\includegraphics[width=\hsize]{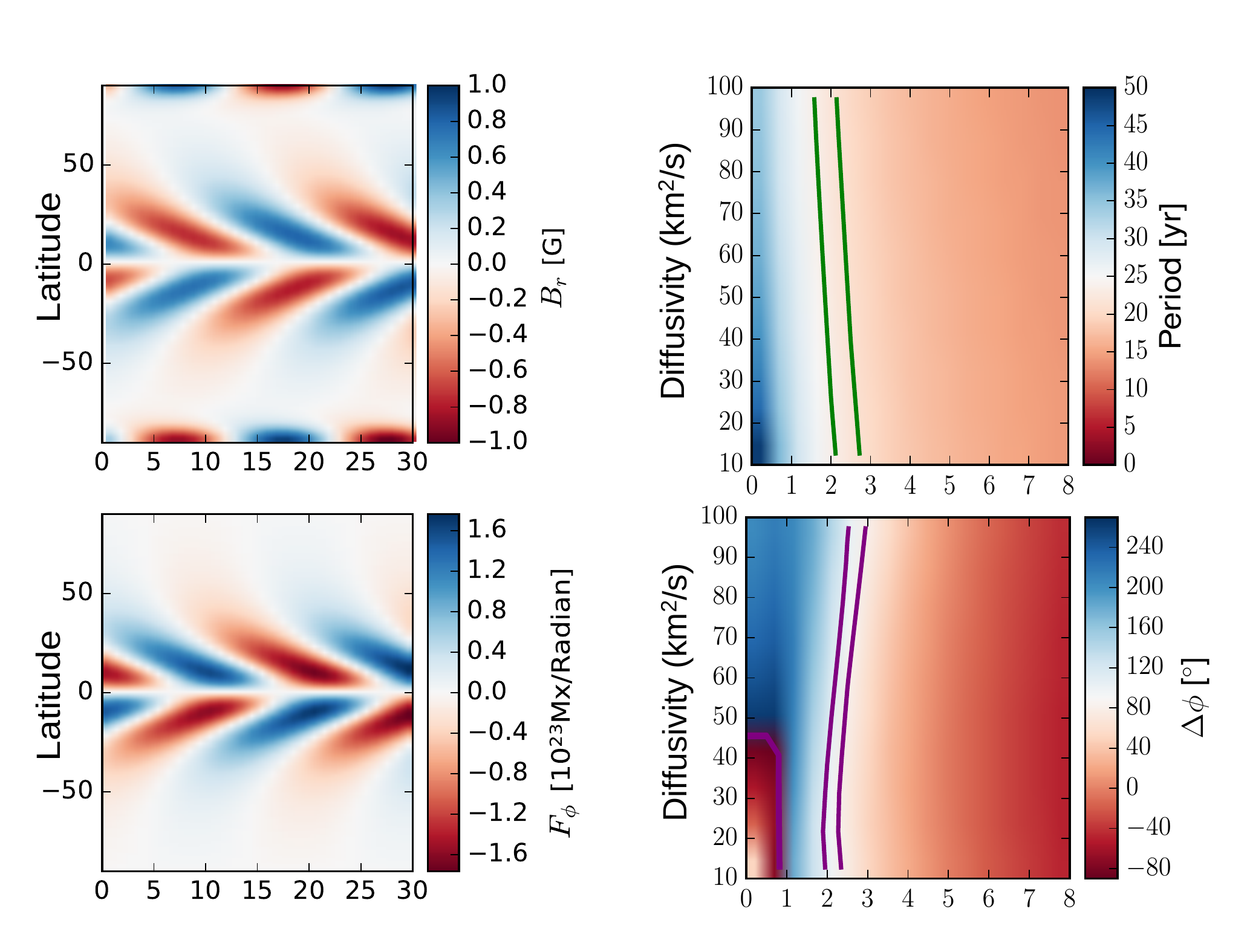}
\caption{Results from the updated Babcock-Leighton model of
  \citet{CameronSchussler::2017b}. {\it Left panels:} Radial surface
  field (upper panel) and depth-integrated toroidal flux (lower panel)
  as a function of latitude and time for a solution consistent with
  the properties of the solar cycle. {\it Right panels:} Properties of
  model solutions as functions of the amplitude of the effective speed
  of the equatorward return flow and of the magnetic diffusivity in
  the convection zone. Colour shadings indicate the dynamo period
  (upper panel) and the phase difference between the maximum of the
  polar field and the maximum rate of flux emergence (lower
  panel). The regions between the coloured lines indicate
  ``solar-like'' models: period in the range 21--23~yr and phase
  difference between 80 and 100 degrees. These results indicate a
  rather tight constraint for the speed of the return flow of about
  2--3$\,$m$\,$s$^{-1}$.}
\label{fig:CS2017b}
\end{figure}

Typically, FT/BL dynamo models are 2D (axisymmetric) or even 3D and
time-dependent.  Although they exhibit solar-like solutions for
properly chosen values of the parameters (e.g., turbulent diffusivity,
geometry and speed of the meridional flow, dynamo excitation, etc.),
computational limitations do not permit a complete coverage of the
parameter space. In order to systematically investigate the parameter
ranges for solar-like solutions, \citet{CameronSchussler::2017b}
updated the 1D, two-layer model of \citet{Leighton:1969}. Model
quantities are the azimuthally averaged radial field at the surface
and the radially integrated toroidal flux, both as a function of
latitude. Poleward meridional flow at the surface, an equatorward
return flow somewhere in the convection zone, latitudinal differential
rotation as well as radial differential rotation in the near-surface
shear layer are included. Furthermore, downward convective pumping of
the near-surface horizontal field and turbulent diffusion of the
radial surface field and the toroidal field are considered. Using the
radially integrated magnetic flux has the advantage that the model
neither depends on the radial distribution of the toroidal field in
the convection zone nor on the depth location of the meridional return
flow.  The solutions of this linear model depend on four parameters
which represent the driving of the dynamo (by emergence of tilted
bipolar regions), the effective speed of the meridional return flow,
the turbulent diffusivity affecting the toroidal field in the
convection zone, and the effective radial shear below the near-surface
shear layer. All other parameters (such as meridional flow and
diffusivity at the surface) are taken from observations. The
simplicity of the model permits the exploration of the full
four-dimensional parameter space. Relevant ranges of parameters are
identified by requiring that the model results meet observational
constraints: positive growth rate, dipole parity, 22-year magnetic
period, flux emergence concentrated in low latitudes, and a 90~deg
phase difference between the maximum of flux emergence and maximum
strength of the polar field. It turns out that these requirements
strongly constrain the parameters, yielding about $2\,$m$\,$s$^{-1}$
for the speed of the return flow, a turbulent diffusivity of about
80$\,$km$^{2}$s$^{-1}$, toroidal flux generation dominated by
latitudinal differential rotation, and weak dynamo excitation not far
above the threshold. A sketch of the model is shown in
Fig.~\ref{fig:CS2017a} and an example of the results in
Fig.~\ref{fig:CS2017b}.

The model results of \citet{Leighton:1969} indicate that the solar
dynamo most probably is a flux-transport dynamo operating near
marginal excitation. The model also clearly favors diffusion-dominated
dynamo action \citep[][]{Yeates:etal:2008}.  The inferred speed of the
equatorward flow transporting the toroidal flux is consistent with the
latitudinal drift rate of the activity belts.  The relevant values of
the magnetic diffusivity affecting the toroidal flux are also
consistent with the observed evolution of the solar activity belts
\citep{Cameron:Schuessler:2016}.  Dominance of latitudinal
differential rotation entails a natural explanation for the
concentration of flux emergence in low latitudes: the latitudinal
rotational shear peaks at mid latitudes and the deep meridional return
flow transports the generated toroidal flux equatorward. Consequently,
it is not necessary to assume a threshold value of the toroidal field
for the initiation of flux emergence, e.g., in the sense of a buoyancy
instability.

\begin{figure}
\centering
\includegraphics[width=\hsize]{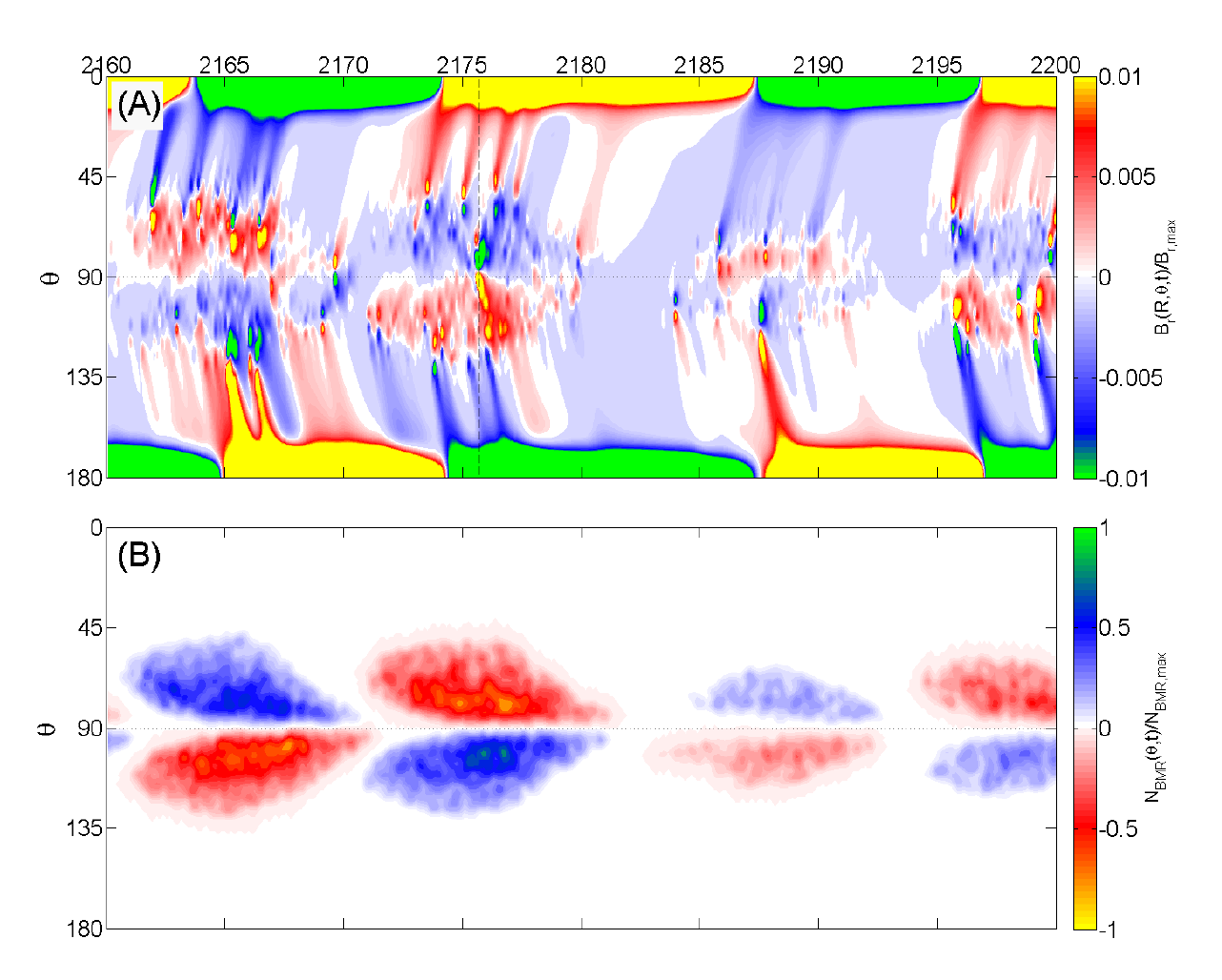}
\caption{Example result obtained by \citet[][part of their
    Fig.~1]{NagyLemerle::2017} using the 2D$\times$2D dynamo model
  developed by \citet{Lemerle:Charbonneau:2017}. Shown are
  time-latitude diagrams of the radial field at the surface (upper
  panel) and of the number of flux emergences (butterly diagram, lower
  panel).}
\label{fig:Nagy}
\end{figure}

Further development of BL dynamo models was provided through a closer
connection between the near-surface evolution and the interior
evolution of the magnetic field. \citet{Cameron:etal:2012} found that
downward convective pumping (leading to predominantly radial field near
the surface) is required to bring a 1D SFT model into accordance with
the results of a 2D axisymmetric FTD model \citep[see
  also][]{Karak:Cameron:2016}.  \citet{Bhowmik:Nandy:2018} used the
surface distribution of magnetic flux during activity minima resulting
from a data-driven SFT simulation as input for a FTD model.  A fully
coupled 2D$\times$2D model was developed by \citet{Lemerle:etal:2015}
and \citet{Lemerle:Charbonneau:2017}. These authors connected an
observationally calibrated SFT model in latitude and longitude with an
axisymmetric FTD model operating in the meridional (radius-latitude)
plane. The FTD model provides the source for the SFT model by
stochastic flux emergence in tilted bipolar magnetic regions. In turn,
the SFT model feeds back upon the FTD model in the form of a BL-like
source term near the upper boundary of the dynamo model, resulting in
a self-consistent coupling of the two models. Fig.~\ref{fig:Nagy}
illustrates that properly calibrated results from this model are
consistent with the characteristic features of the corresponding solar
observations. A similar 2D$\times$2D approach was presented by
\citet{Miesch:Dikpati:2014}.

The BL source in such models is typically provided by ``deposition''
of tilted bipolar regions in the SFT part of the model, non-locally
depending on the strength of the deep-seated toroidal field. This
ad-hoc procedure ignores the details of the formation and rise of flux
loops through the convection zone, in particular the development of
the tilt angle. It also implicitely assumes a dynamical disconnection
of the emerged bipolar region from its subsurface roots.
\citet{Bekki:Cameron:2022} considered the post-emergence evolution of
deposited bipolar regions in the framework of a nonlinear 3D
simulation in a spherical shell, including the Lorentz force as well
as solar-like differential rotation and meridional flow driven by a
mean-field prescription.  These authors found that the evolution
depends sensitively on the initial shape and depth of the injected
bipolar regions. When initialized with zero tilt, the Lorentz force in
combination with the Coriolis force tends to produce negative tilt
angles. The simulated BMRs also develop an systematic asymmetry
between the strength of leading and following spots, which is
consistent with observations.

\citet{Yeates:Munoz-Jaramillo:2013} suggested a more consistent
treatment of flux emergence by assuming helical upflows that transport
flux loops through the convection zone, thus capturing aspects of
buoyancy and advection by convective flows as well as the connection
between the surface and interior field. A similar treatment was
proposed by \citet{Pipin:2022}. The idea of this approach is
consistent with the results of detailed observations of the properties
of emerging flux, which favor passive flux transport by convective
upflows \citep{Birch:etal:2016}. On the other hand, comparison with
well-calibrated SFT models \citep{Whitbread:etal:2019} shows that
after emergence the surface flux needs to be dynamically disconnected
from its deep toroidal roots as suggested by
\citet{Schuessler:Rempel:2005}. Moreover, \citet{Schunker:etal:2019}
and \citet{Schunker:etal:2020} showed that the tilt of bipolar
magnetic regions develops only {\it after} emergence, possibly related
to the extended flows converging towards the bipolar magnetic regions
\citep{Martin-Belda:Cameron:2016, Gottschling:etal:2021}.  In view of
these results it seems fair to say that the formation of rising flux
loops, their transfer through the convection zone, the actual
emergence process, and the subsequent early evolution of the emerged
magnetic flux are still poorly understood \citep[see also reviews
  by][]{Fan:2021, Isik:etal:2023}, so that these processes cause a
significant amount of uncertainty in all FTD/BL models presented so
far.

Among the first 3D studies of FTD/BL models for the solar dynamo are
the models of \citet{Hazra:etal:2017} and
\citet{Karak:Miesch:2017}. While these authors used the deposition
approach to represent flux emergence, \citet{Kumar:etal:2019}
implemented the more consistent recipe of
\citet{Yeates:Munoz-Jaramillo:2013} in a kinematic 3D dynamo code.
Another step towards a full 3D treatment was taken by
\citet{Hazra:Miesch:2018}, who introduced a 3D, stationary,
convection-like flow field in the upper part of the convection
zone. This serves to replace the diffusion term in standard SFT models
by a more realistic explicit transport process.  The assumed flow
pattern was based upon an observational power spectrum of the surface
flows and a downward extrapolation under the assumption of zero
divergence of the mass flux and an imposed radial profile. The
kinematic model ran into problems owing to unlimited small-scale
dynamo action driven by the imposed stationary flow. Further results
obtained with kinematic 3D models of BL dynamos have been reviewed by
\citet{Hazra:2021}.  \citet{Bekki:Cameron:2022} used their 3D model to
perform nonlinear BL dynamo simulations with the near-surface
deposition of properly tilted bipolar regions. The Lorentz force
provides nonlinear saturation of the dynamo amplitude. This feedback
drives non-axisymmetric flows as well as solar-like zonal flows
superposed upon the differential rotation and also modifies the
meridional circulation.  Other nonlinear effects (see also Sect.
\ref {subsec:nonlin} that could lead to dynamo saturation and
determine the cycle amplitude have been studied by \citet{Jiang:2020},
\citet{Jiao:etal:2021}, and \citet{Talafha:etal:2022}.

\section{Generation and removal of magnetic flux}\label{sec:gen}

\subsection{Toroidal flux}\label{subsec:btor}

\begin{figure}
\centering
\includegraphics[width=85mm]{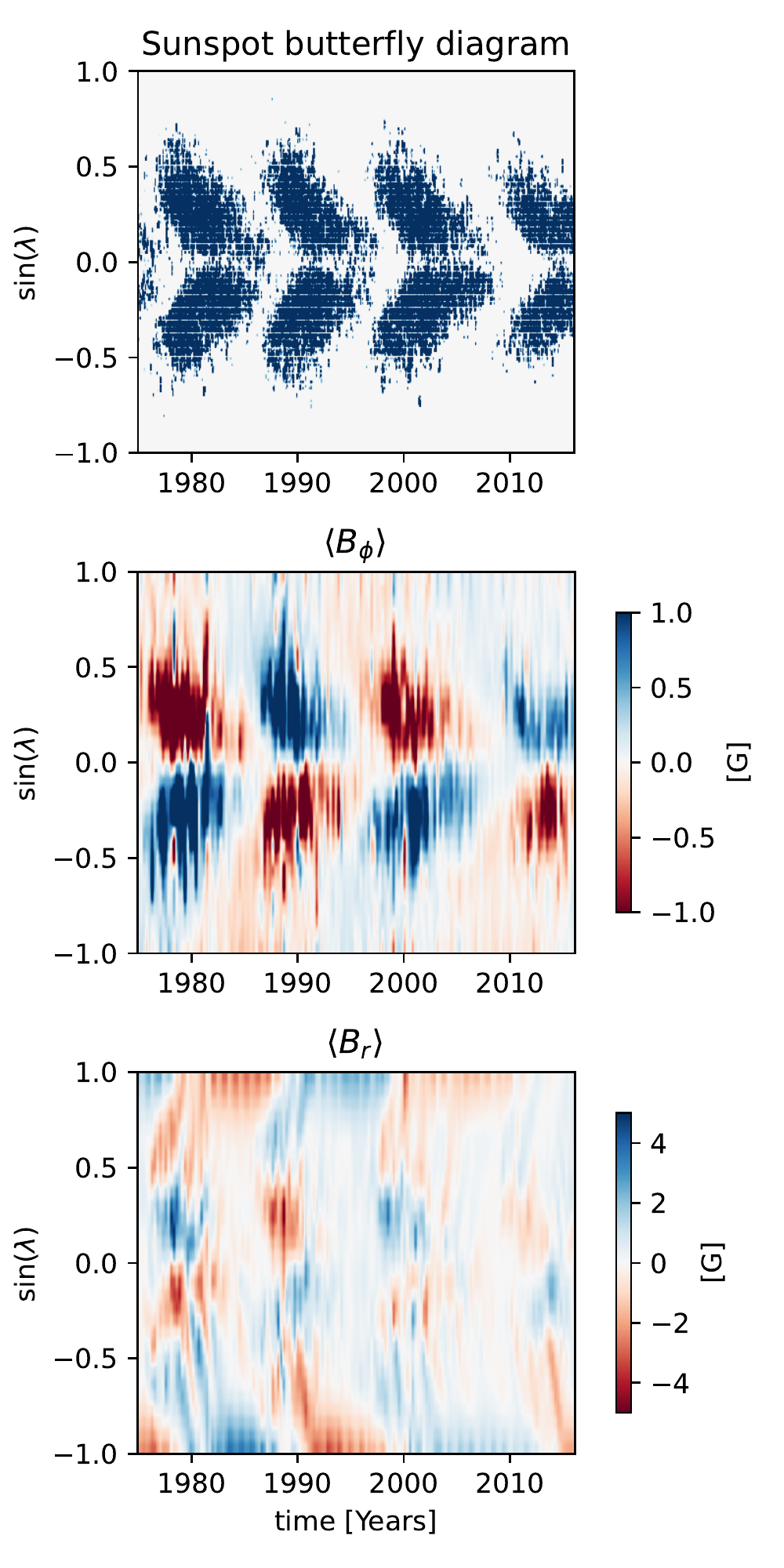}
\caption{Observationally based time-latitude diagrams of sunspots (top
  panel, data from Royal Greenwich Observatory/NOAA:
  http://solarcyclescience.com/AR\_Database), longitudinally averaged
  azimuthal surface field (middle panel, data from Wilcox Solar
  Observatory, WSO), and longitudinally averaged radial field (lower
  panel, data from WSO).}
\label{Fig:bfs}
\end{figure}

Toroidal magnetic field (azimuthal field in the case of an average
over longitude) is considered to be generated in the Sun through the
action of differential rotation on a poloidal magnetic field. Part of
the generated toroidal flux emerges when radial motions carry loops of
magnetic flux through the solar surface. Flux emergence leads to the
formation of two surface areas with oppositely-directed radial field
components. These bipolar magnetic regions (BMRs) have been studied
extensively \citep[e.g.][]{HaleEllerman::1919, HarveyHarvey::1975,
  MartinHarvey::1979, WilsonAltrocki::1988,Harvey::1993,
  McClintockNorton::2016, SchunkerBraun::2016}. A map showing the
latitudes at which larger BMRs with sunspots emerge in time (known as
the butterfly diagram) is shown in the upper panel of
Fig.~\ref{Fig:bfs}.

BMRs show various systematic properties.  One example is Hale's law
\citep{HaleEllerman::1919}, which states that the leading polarity
(with respect to the direction of rotation) of BMRs in each hemisphere
has the same sign in most (up to 95\%) of the cases during an activity
cycle. The leading polarity is opposite in the two hemispheres and
switches from cycle to cycle. These properties imply a systematic
East-West component of the horizontal field during emergence, which
flips sign between each hemisphere and from cycle to cycle. This can
be illustrated in a time-latitude map of the longitudinally averaged
azimuthal component of the surface field covering four activity
cycles, which is shown in the middle panel of Fig.~\ref{Fig:bfs}
\citep[cf.][]{CameronDuvall::2018, LiuScherrer::2022}.  The latitudes
where sunspot groups emerge correspond to locations where the surface
azimuthal field is strong. Weaker azimuthal field corresponds to flux
emergence of smaller BMRs \citep[ephemeral regions,][]{MartinHarvey::1979}.

Another systematic property of BMRs is Joy's law
\citep{HaleEllerman::1919}. It refers to the tendency for the leading
part of a BMR or sunspot group to be closer to the equator. Because of
Hale's law, the leading parts mostly have the same polarity, so that
there is a systematic tendency in both hemispheres that more radial
magnetic flux of one polarity appears closer to the equator.  The
radial field of the leading parts in the other hemisphere has the
opposite polarity, and the polarities switch from cycle to cycle.
These properties are illustrated in the time-latitude diagram for the
longitudinally averaged radial surface field shown in the bottom panel
of Fig.~\ref{Fig:bfs}.  The locations of flux emergence can be clearly
seen, with the leading polarity dominating on the equatorward side of
the ``butterfly wings''. Plumes of magnetic field with widths of a few
months connect the butterfly wings to the polar regions, where the
field switches polarity with the period of the solar cycle.

\begin{figure}
\includegraphics[width=\textwidth]{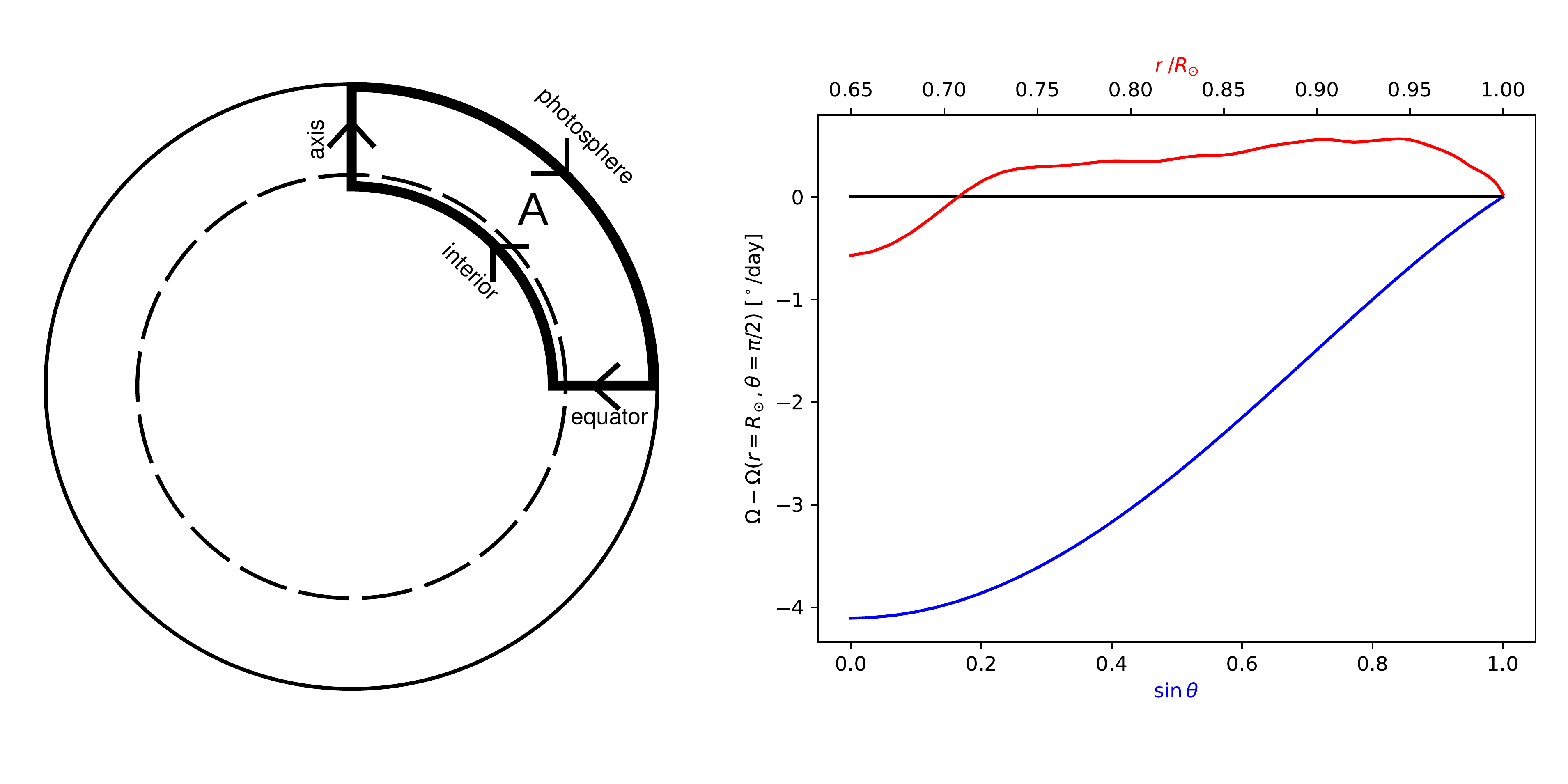}
\caption{{\sl Left panel:} Area and contour used for calculating the
  net axisymmetric toroidal flux in a hemisphere. The bottom part of
  the contour is placed at a radius $R_{\mathrm{interior}}$, where the
  penetration of the 22~year cyclic component of the magnetic field is
  negligible.  {\sl Right panel:} Solar differential rotation along
  segments of the contour in the left panel. Shown are the radial
  dependence of $\Omega$ on the equatorial plane (red curve, upper
  scale) and the (co)latitudinal dependence at the surface (blue
  curve, lower scale), both relative to the equatorial surface
  rotation rate.  The data for this plot was obtained from
  \cite{LarsonSchou::2018}.}
\label{fig:contour+DR}
\end{figure}

An important feature of the time-latiude diagram of the toroidal
surface field (middle panel of Fig.~\ref{Fig:bfs}) is that during
times of activity maximum (when there are many sunspots) the toroidal
field at every latitude in each hemisphere is of the same sign.  This
suggests to consider the evolution of the net subsurface toroidal
flux in each hemisphere. For this we start from the induction equation,
\begin{equation}
  \frac{\partial{\bf B}}{\partial t}=\nabla \times \left( {\bf U} \times {\bf B}  - \eta \nabla \times {\bf B} \right),
\end{equation}
where ${\bf B}$ is the magnetic field, ${\bf U}$ the flow velocity,
and $\eta$ the (molecular) magnetic diffusivity. Following
\citet{Cameron:Schuessler:2015}, \citet{CameronSchuessler::2020}, and
\citet{JeffersCameron::2022}, we now consider the area A that is
enclosed by the thick outline in the left panel of
Fig.~\ref{fig:contour+DR} and apply Stokes' theorem to the induction
equation, viz.
\begin{equation}
  \frac{\partial{\iint_A {\bf B} \cdot \mathrm{d}{\bf A}}}{\partial t}=\oint_{\delta A} \left( {\bf U} \times {\bf B}  - \eta \nabla \times {\bf B} \right) \cdot \mathrm{d} {\bf l} \,,
\end{equation}
\label{eq:stokes} 
where $\delta A$ indicates the boundary of $A$ and ${\mathrm{d}\bf l}$
is the corresponding line element.  We now take the azimuthal average of
Eq.~(2) and neglect the diffusive term in the contour
integral since the magnetic Reynolds number is very large. This leads
to
\begin{equation}
  \frac{\partial{\iint_A {\langle {\bf B} \rangle} \cdot \mathrm{d}{\bf A}}}{\partial t}=\oint_{\delta A} \left( \langle{\bf U}\rangle \times \langle{\bf B}\rangle
+ \langle{\bf u}\times {\bf b}\rangle \right) \cdot \mathrm{d} {\bf l} \,,
\end{equation}
where $\langle\dots\rangle$ indicates the azimuthal average. This
equation represents the temporal change of net toroidal flux in the
Northern hemisphere resulting from the inductive action of the flow
field on the magnetic field.  The quantities ${\bf b}={\bf B}-\langle
{\bf B} \rangle$ and ${\bf u}={\bf U}-\langle {\bf U} \rangle$ denote,
respectively, the fluctuating components of magnetic field and flow
velocity with respect to the corresponding azimuthal averages.

Since the Sun's radial differential rotation in the equatorial plane
is small compared to its latitudinal differential rotation (see right
panel of Fig.~\ref{fig:contour+DR}) it is convenient to use a frame of
reference rotating with the surface equatorial rate. In
this frame, we have
\begin{eqnarray}
  \oint_{\delta A} \left( \langle{\bf U}\rangle \times \langle{\bf B}\rangle \right) \cdot \mathrm{d} {\bf l} &=&
  \int_0^{\pi/2} \left( \Omega(R_\odot,\theta)-\Omega(R_\odot,\frac{\pi}{2})\right) \sin\theta \langle{B_r}\rangle R_\odot^2 \mathrm{d} \theta \nonumber \\
  & & -\int_{R_\mathrm{interior}}^{R_\odot} \left(\Omega(r,\frac{\pi}{2})-\Omega(R_\odot,\frac{\pi}{2})\right) \langle{B_\theta}\rangle  r \mathrm{d}r \,,
\end{eqnarray}
where $\theta$ is the colatitude. Only the surface part of the contour
integral (first term on right-hand side) and the part in the
equatorial plane (second term) contribute to the contour integral. The
bottom part in the interior vanishes since {\bf B} is zero there and
the part along the rotation axis vanishes since we consider azimuthal
averages of all quantities.  The solenoidality of the magnetic field
implies
\begin{equation}
\int_0^{\pi/2}  \langle{B_r}\rangle R_\odot^2 \sin\theta \mathrm{d} \theta 
  = -\int_{R_\mathrm{interior}}^{R_\odot} \langle{B_\theta}\rangle  r \mathrm{d}r,
\end{equation}
which means that the net magnetic flux through the equatorial plane
beneath the surface is equal in magnitude to the net flux through the
solar surface in each hemisphere.

Owing to the weak radial differential rotation in the equatorial
plan), the net toroidal flux generation is strongly dominated by the
surface part of the contour integral, i.e., the first term on the
right-hand side of Eq.~(4). This can be directly evaluated from
synoptic observations of the radial field on the surface and the Sun's
differential rotation, with a net hemispheric toroidal flux of $5$--$8
\times 10^{23}$~Mx per cycle being thus generated. The integrand for
the surface part in the contour integral (Eq.~4) is strongly dominated
by the polar caps (see bottom left panel of Fig.~6), so that the flux
associated with the polar dipole field represents the relevant
poloidal source for the generation of net toroidal flux.  Other
poloidal flux that is contained in the convection zone and does not
cross the surface leads to equal amounts of East-West and West-East
orientated toroidal flux in a hemisphere and thus does not contribute
to the net toroidal flux required by Hale's polarity laws.

The term in Eq.~3 involving the fluctuating components, $\oint_{\delta
  A} \langle{\bf u}\times {\bf b}\rangle \cdot \mathrm{d} {\bf l}$, is
dominated by (turbulent) diffusive fluxes across the axis and the
equatorial plane.  Some of the other terms, such as those owing to the
turbulent electromotoric force ($\alpha$-effect), are expected to
vanish due to symmetries -- along the axis owing to the conservation
of helicity and at the equator owing to the expected vanishing of the
kinetic helicity, which in its simplest form scales like $\cos\theta$.

Another contribution from the fluctuating terms is the change of the
net toroidal flux by flux emergence and submergence through the
surface, which is described by 
\begin{equation}
  \left(\frac{\mathrm{d} \Phi}{\mathrm{d} t}\right)_{\rm em,subm} 
   =\int^0_{\pi/2} \langle u_r b_\phi \rangle\vert_{R_\odot} R_{\odot}  
   \mathrm{d}\theta\,. 
\label{eq:em_subm}
\end{equation}
Flux loss through the surface was part of the original
Babcock-Leighton model \citep{Leighton:1969}, but later fell out of
favour since it was thought that the emerged flux is mostly retracted
back through the surface \citep[e.g.,][]{Wallenhorst:Topka:1982}. This
process can take place in flux cancellation events when loops of
emerged flux become sufficiently narrow, so that the tension force
dominates. Some retraction of flux is possibly required to account for
the amount of flux brought to the surface by repeated emergences in
so-called ``nests of activity'' \citep[e.g.,][]{Gaizauskas:etal:1983}.

\citet{Parker::1984} argued that a net loss of toroidal flux from the
Sun would require an organized sequence of reconnection events between
narrowly-spaced bipolar regions, thus effectively shedding the mass
from the toroidal field lines and allowing them to freely escape with
the solar wind. However, observations show that the bipolar regions
generally are much too widely spaced for this process to operate on
the Sun, so that Parker concluded that in dynamo models the solar
photosphere should be approximated by an impenetrable boundary.

Parker's argument, however, ignores that what is relevant for the
dynamo models is the effect of the emergence on the {\it mean\/}
(azimuthally averaged) toroidal field. \citet{CameronSchuessler::2020}
showed that the amount of the azimuthally averaged toroidal field lost
due to the emergence of a bipolar region is time-independent,
irrespective of the further evolution of its magnetic flux.  While
cancellation/retraction of most of the field may occur, at the same
time the remaining surface flux spreads out in longitude (transported
by near-surface flows) to eventually fully encircle the Sun. In this
way, toroidal field detached from the solar interior is formed which
can be carrried away by the solar wind and the same amount of flux is
lost from the mean toroidal field in the interior.  A quantitative
analysis of the net azimuthally averaged toroidal flux lost during a
cycle arrives at $3.3 \times 10^{23}$~Mx/hemisphere/cycle
\citep{CameronSchuessler::2020}. A similar estimate of $5\times
10^{23}$ was obtained using in~situ measurements of the magnetic field
and flows in the solar wind \citep{BieberRust::1995}.  The
observations thus indicate that, integrated over a cycle, the amount
of the toroidal magnetic flux generated by the axisymmetric flows and
fields ($\oint_{\delta A} \left( \langle{\bf U}\rangle \times
\langle{\bf B}\rangle \right) \cdot \mathrm{d} {\bf l}$) is similar to
the amount lost through flux emergence ($\oint_{\delta A} \langle{\bf
  u}\times {\bf b}\rangle \cdot \mathrm{d} {\bf l}$).

\begin{figure}
  \includegraphics[width=\textwidth]{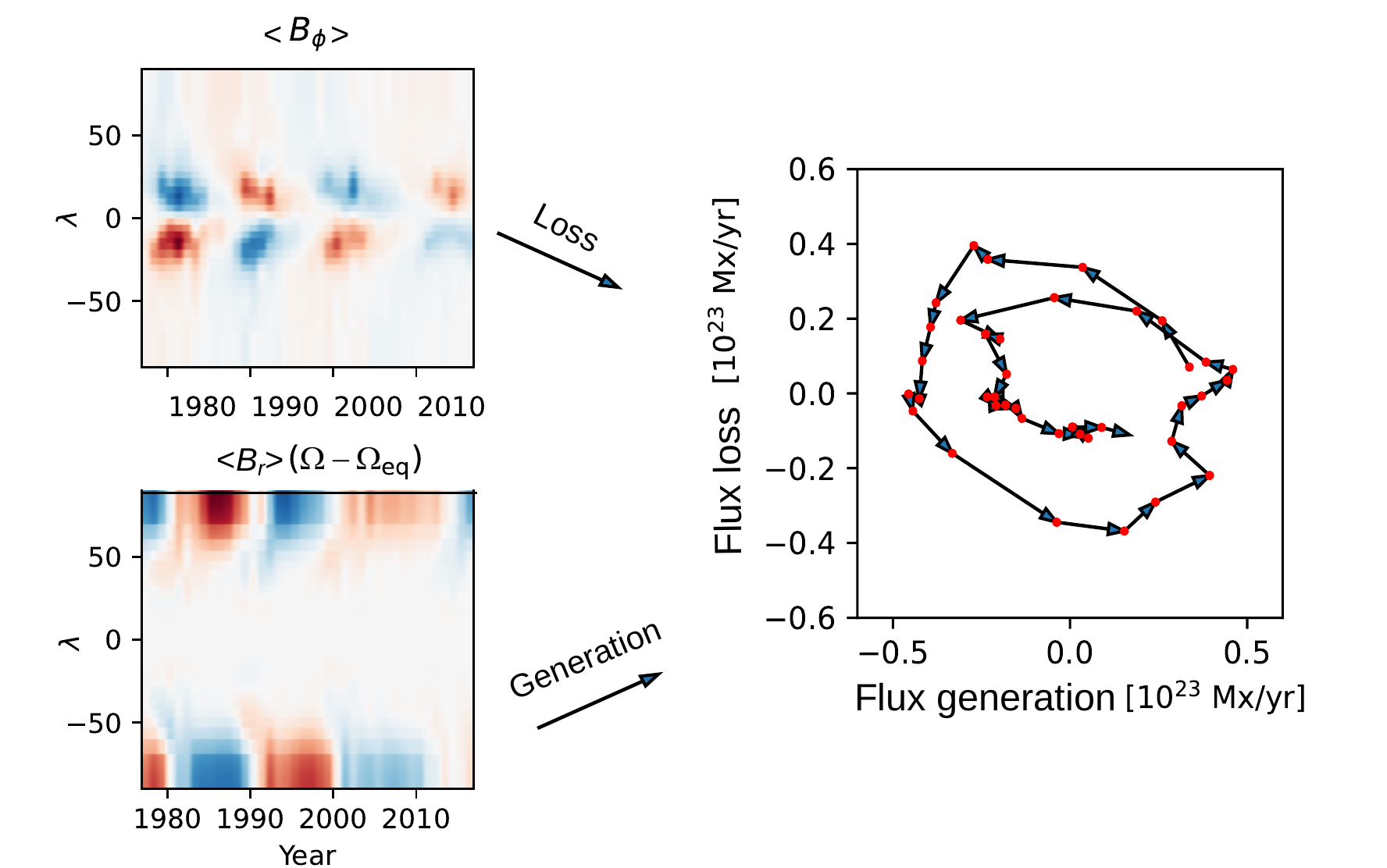}
  \caption{Generation and loss of net unsigned toroidal flux in the
    Northern hemisphere, covering four activity cycles on the basis of
    synoptic magnetograms from the Wilcox Solar Observatory.  The
    time-latitude diagrams on the left side show the observational
    input for evaluating the dominating surface part of the contour
    integral (Eqs.~4 and 6). Toroidal flux is generated proportional
    to the product of the azimuthally averaged surface radial field
    and the surface rotation rate relative to the equatorial rate
    (lower left panel).  Toroidal flux is lost through flux emergence
    at a rate proportional to the product of $\langle B_\phi \rangle$
    at the surface (upper left panel) and the radial emergence
    velocity estimated as $v_\mathrm{em} = 200\,$m$\,$s$^{-1}$
    \citep{Centeno:2012}. The panel to the right shows a phase plot of
    the yearly integrated values of the generated and lost toroidal
    flux indicated by red dots. The connecting line segments with
    arrows illustrate the change from one year to the next.}
\label{fig:toroidal_balance}
\end{figure}

The toroidal flux budget shown in Figure~\ref{fig:toroidal_balance}
shows the approximate balance between the amount of toroidal magnetic
flux lost throughout a cycle due to flux emergence and the amount
produced by the winding up of poloidal flux threading the solar
surface, both being determined by surface observations of the magnetic
field ($B_\phi$ and $B_r$).  There is a phase shift between the
production rate and loss of about 90$^\circ$, which is consistent with
the loss through flux emergence being proportional to the total
subsurface toroidal flux. The phase diagram is based on Eqs.~(4) and
(6), which confirms the concept that the toroidal flux is mainly
generated by the action of latitudinal differential rotation on the
poloidal flux threading the solar surface in the polar dipole field.

\subsection{Poloidal flux}\label{subsec:bpol}

The previous section has shown that the application of Stokes' theorem
to the induction equation over a properly defined meridional area
illuminates the essence of the generation and loss of net toroidal
flux during solar activity cycles. A similar procedure can be used to
determine the evolution of the net radial flux threading a hemisphere
of the Sun \citep[cf.][]{DurrantTurner::2004}.  To this end, we
integrate the radial component of the azimuthally averaged induction
equation over the photospheric surface area, $A$, of the Northern
hemisphere, NH, and apply Stokes' theorem, which yields
\begin{equation}
  \frac{\partial\Phi_{\mathrm{pol, NH}}}{\partial t} = 
  \frac{\partial}{\partial t}
  \oint_{A} B_r dA = 2 \pi R_\odot \left( \langle u_r b_\theta \rangle - \langle u_\theta b_r \rangle \right) \,, 
\end{equation}
where the contour integral is taken over the boundary of $A$,
i.e., the equator.  The two terms on the r.h.s. correspond to
observable quantities. The term $\langle u_r b_\theta \rangle$
quantifies the effect of flux emerging across the equator. This term
is important when large and highly tilted active regions emerge with
polarities on either side of the equator
\citep[e.g.,][]{CameronDasi-Espuig::2013}.  The term $\langle u_\theta
b_r \rangle$ corresponds to convective motions carrying
radial flux across the equator and can be approximated by a
``turbulent'' diffusion term, viz.
\begin{equation}
  \langle u_\theta b_r \rangle= \eta_{\mathrm{T}} R_\odot
  \frac{\partial B_r}{\partial \theta} \,.
\end{equation}  
There is a systematic component to ${\partial B_r}/{\partial \theta}$
corresponding to the systematic tilt angle of bipolar regions (see the
time-latitude diagram of the radial magnetic field in
Fig.~\ref{Fig:bfs}). A random component is introduced by large, highly
tilted active regions which emerge near to the equator
\citep{CameronDasi-Espuig::2013}, also called ``rogue active
regions'' \citep{NagyLemerle::2017}.  Such regions introduce
significant randomness into the net flux in each hemisphere and the
axial dipole moment \citep{CameronJiang::2014}. This randomness thus
carries over into the production of the toroidal flux emerging in the
subsequent cycle. \cite{JiangCameron::2015} and
\cite{WhitbreadYeates::2018} showed that the low amplitude of solar
cycle 24 can be understood in terms of rogue active regions that
emerged during cycle 23.

\begin{figure}
  \includegraphics[width=\textwidth]{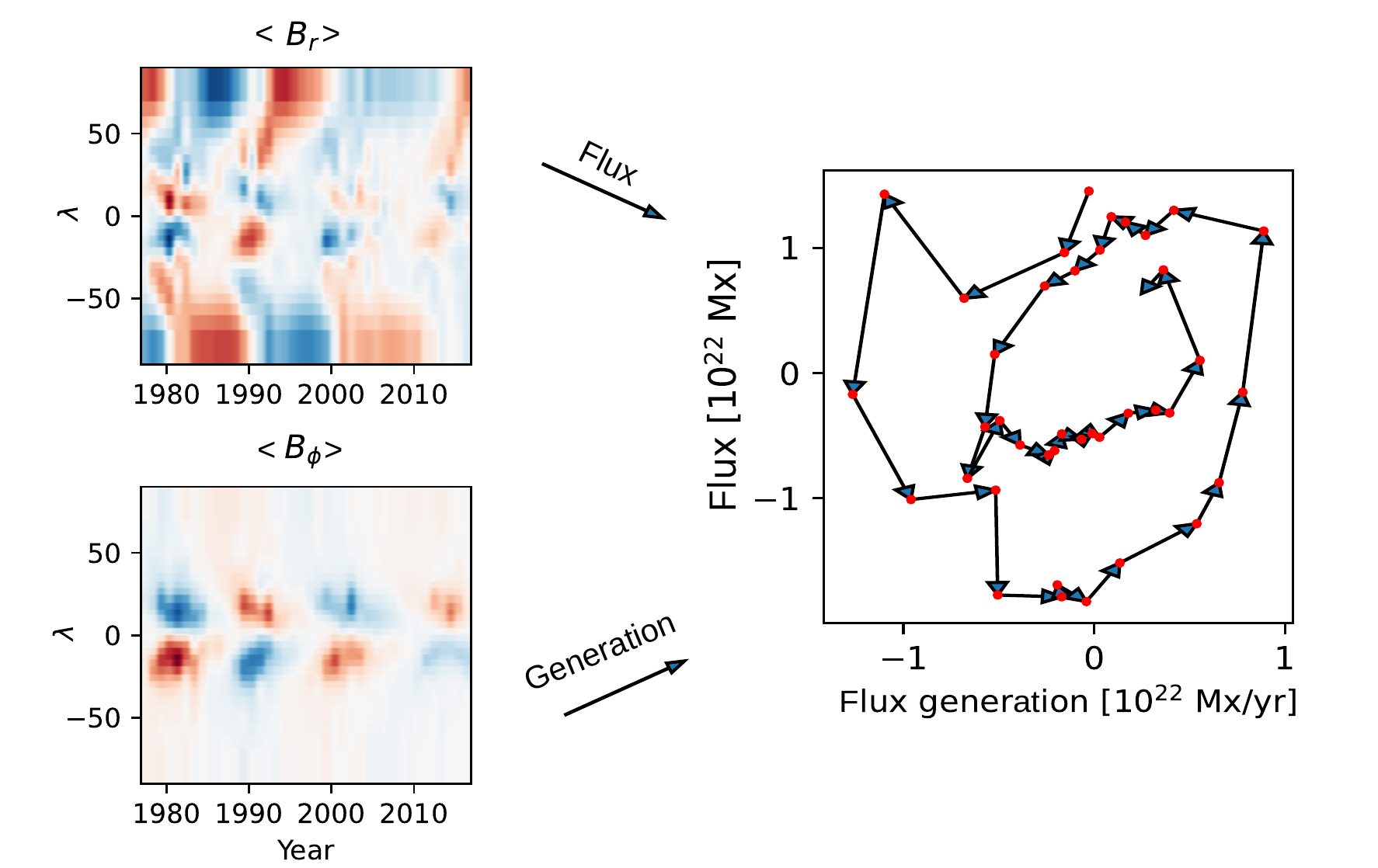}
  \caption{Systematic part of the generation rate and amount of net
    poloidal flux in the course of four activity cycles.  The left
    panels illustrate the observational input: the azimuthally
    averaged radial surface field (upper left panel) determines the
    total net poloidal surface flux (cf. Eq.~9) while the
    azimuthally averaged azimuthal surface field (lower left panel)
    represents flux emergence in tilted bipolar regions and thus
    determines the rate of generation of poloidal flux via flux
    transported over the equator (cf. Eq.~15).  The right panel shows
    a phase plot of both quantities, with each red dot representing the the
    average value for one year. The line segments with arrows
    represent the change from one year to the next.}
\label{fig:poloidal_balance}
\end{figure}

Figure~\ref{fig:poloidal_balance} illustrates the budget of poloidal
surface flux for four activity cycles on the basis of synoptic
magnetograms from the Wilcox Solar Observatory. The poloidal flux is
determined by integration of the longitudinally averaged radial field
over the solar surface, viz.
\begin{equation}
\Phi_{\mathrm{pol}}= \pi R_\odot^2 \int_{0}^{\pi} \langle{B_r}\rangle
                    \mathrm{sign}(\pi/2-\theta) 
                    \sin\theta \mathrm{d}\theta \,,
\end{equation}
where the results for the two hemispheres are averaged.

The generation term for the poloidal flux results from summing the
contributions of all bipolar regions to the amount of magnetic flux
which is carried across the equator.  The contribution of a single
bipolar region ($i$) depends on the magnetic flux of each
polarity, $\Phi_{i}$, the angular separation of the two polarities in
latitude, $\delta_{\lambda,i}$, the latitude of emergence, $\lambda_{i}$,
and a parameter characterizing the surface evolution of the magnetic
flux near the equator, $\lambda_R$
\citep[cf.][]{PetrovayNagy::2020}. This parameter depends on the
latitudinal derivative of the surface meridional flow speed, $V$, at
the equator ($\lambda=0$), and the turbulent diffusivity at the
surface, $\eta_{\mathrm T}$, viz.
\begin{equation}
  \lambda_R=\sqrt{\frac{\eta_{\mathrm T}}{R_\odot 
     {(\mathrm{d}V}/{\mathrm{d}\lambda})_{\lambda=0}}} \,.
\end{equation}
\citet{PetrovayNagy::2020} estimated the time-integrated amount of
magnetic flux crossing the equator (and thus the contribution of the
bipolar region to the buildup of the poloidal field) by the leading
term of a Taylor expansion as
\begin{equation}
 \Phi_{{\mathrm{pol}},i} = \frac{\Phi_{i} \delta_{\lambda,i}}{\sqrt{2\pi}\lambda_R}  
    \; {\mathrm e}^{-\lambda^2_{i}/ (2 \lambda_R^2)} \,. 
\label{eq:petrovay} 
\end{equation}
During its emergence, the flux tube forming the bipolar region
contributes to the longitudinally averaged toroidal surface field,
$\langle B_{\phi} \rangle$, by the amount $\langle B_{\phi,i} \rangle$
according to
\begin{equation}
  \Phi_{i} \frac{\delta_{\phi,i}}{2\pi}=\langle B_{\phi,i} \rangle
  v_{\mathrm{em}}\, \Delta t \,R_\odot\, \Delta\lambda \,,
\end{equation}
where $\delta_{\phi,i}$ is the longitudinal angular extent of the
bipolar region, $\Delta t$ is the time over which the emergence takes
place, $\Delta\lambda$ is the latitudinal width of the leading
polarity of the emerging region, and $v_{\mathrm{em}}$ is the mean
emergence speed (assumed to be the same for all emergences).  The
latitudinal and longitudinal extents are related by
$\delta_{\lambda,i}= \tan(\alpha_i)\cos(\lambda_i)\delta_{\phi,i}$,
where $\alpha_i$ is tilt angle.  We thus obtain
\begin{equation}
  \Phi_{i} \delta_{\lambda,i}=\langle B_{\phi,i} \rangle \,
  v_{\mathrm{em}}\, \Delta t \, R_\odot \, \Delta\lambda \, 
  2\pi \tan(\alpha_i)\cos(\lambda_i)  \,.
\end{equation}
Inserting this result into Eq.~(\ref{eq:petrovay}), we obtain
\begin{equation}
  \Phi_{{\mathrm{pol}},i} = \frac{\langle B_{\phi,i} \rangle \,
    v_{\mathrm{em},i}\, \Delta t \, R_\odot \, \Delta\lambda \, \sqrt{2\pi}
    \, \tan(\alpha_i) \, \cos(\lambda_i)}{\lambda_R} \,
  \mathrm{e}^{-\lambda^2_{i}/ (2 \lambda_R^2)}
\end{equation}
For the the tilt angle we consider Joy's law without
scatter, so that $\alpha_i=\alpha(\lambda_i)$. We further assume that
the emergence velocity is the same for all bipolar regions with a
value of $v_{\mathrm{em}}=200\,$m$\,$s$^{-1}$ \citep{Centeno:2012}.
Adding together the contributions of the individual bipolar regions to
obtain the rate of generation of poloidal flux then amounts to the latitude
integral
\begin{equation}
\frac{{\mathrm d}\Phi_{\mathrm{pol}}}{\mathrm d t} = 
    \frac{\sqrt{2\pi}\, v_{\mathrm{em}}}{\lambda_R}
    \int_{-\pi/2}^{\pi/2} \langle B_{\phi} \rangle R_\odot 
    \tan[\alpha(\lambda)]\cos(\lambda)\,
    {\mathrm e}^{-\lambda^2/ (2 \lambda_R^2)} {\mathrm d}\lambda
\end{equation}
We choose the parameters as in \citet{JiangCameron::2014} with
$\eta_{\mathrm T}=250$~km$^2$s$^{-1}$, surface meridional flow $V(\lambda)=
11\sin(180\lambda/75)\,$m$\,$s$^{-1}$, and Joy's law for a cycle of
intermediate strength in the form $\alpha(\lambda)=0.7\times1.3
\sqrt{\vert\lambda\vert}\, \mathrm{sign}(\lambda)$.

The poloidal flux budget resulting from using the observed
longitudinally averaged radial and azimuthal surface fields in
Eqs.~(9) and (15) is illustrated by the phase diagram on the right
side of Figure~\ref{fig:poloidal_balance}. It demonstrates that
poloidal flux and poloidal flux generation as determined through
Eqs.~(9) and (15) are consistent with each other. This confirms the
basic concept that the poloidal field is being generated by the joint
contributions of tilted bipolar regions.  Compared to the toroidal
flux budget (Figure~\ref{fig:toroidal_balance}), the phase diagram is
somewhat less smooth. This is a consequence of the significant random
component affecting the generation of poloidal flux. This component is
only included in the poloidal flux but not in the rate of poloidal
flux generation, where we have assumed Joy's law without scatter.

\subsection{Flux transport}\label{subsec:transport}
The net toroidal and poloidal flux budgets discussed in the previous
sections show that the observable magnetic flux at the surface plays a
crucial role in the solar dynamo, highlighting the role of
differential rotation, flux emergence, and Joy's law.  This section
briefly discusses additional processes which modifiy the spatial
distribution of the fluxes.  In terms of the poloidal flux, the budget
shown in Fig.~\ref{fig:poloidal_balance} demonstrates that flux carried
across the equator is essential. Furthermore, the toroidal flux
budget is dominated by the polar fields, which means that the poleward
transport of flux is an important ingredient of the dynamo process:
surface flux which crosses the equator is transported
to the poles by a combination of the large-scale meridional flow and
small-scale flows associated with convection. These processes are
captured by the surface flux transport model
\citep{Jiang:etal:2014, Yeates:etal:2023}.

The butterfly diagram of flux emergence demonstrates that there is a
similar requirement for the equatorward transport of the subsurface
toroidal flux from latitudes of about $55^\circ$, where the toroidal
flux is most efficiently produced \citep[cf.][]{Spruit:2011}, towards the
equator. Low-latitude flux emergence in accordance with Joy's law then
leads to the cross-equatorial transport of poloidal field.  The nature
of this transport of toroidal flux is not directly constrained, so
that dynamo waves, turbulent pumping, and meridional circulation are
all viable candidates.  \cite{GizonCameron::2020} showed that the
speed of the deep equatorward meridional flow inferred from
helioseismology is consistent with the observed migration of the
activity belts if the toroidal flux is distributed over the lower half
of the convection zone. This lends credibility to the concept of flux
transport dynamos \citep[recently reviewed by][]{Hazra:etal:2023}.

\section{Nonlinearity, predictability, and long-term variability}\label{sec:fluct}

\subsection{Nonlinear effects}\label{subsec:nonlin}

An excited hydromagnetic dynamo leads to exponential growth of an
initially weak magnetic field until further amplification becomes
limited by the action of the Lorentz force, which introduces a
nonlinearity into the system. The nonlinearity can affect large-scale
flows (e.g, differential rotation and meridional flow), modify
turbulence effects (e.g., the mean-field $\alpha$-term, turbulent
diffusivity, and turbulent pumping), or change the properties of flux
emergence.  A nonlinearity often invoked in mean-field models is
``$\alpha$-quenching'' \citep{SteenbeckKrause::1969, Stix::1972}, a
parameterization of the decreasing efficiency of turbulence to produce
poloidal field from the toroidal field as the field amplitude
grows. In the case of a system with a high magnetic Reynolds number,
$R_m$, such as the Sun, this nonlinearity is potentially catastrophic
since the ratio of the small-scale field and the mean field scales as
$\sqrt{R_{\mathrm m}}$ \citep{CattaneoVainshtein::1991}.  This causes
saturation of the dynamo at field amplitudes many orders of magnitude
smaller than observed on the Sun. However, catastrophic quenching may
be alleviated by removal of magnetic helicity from the system
\citep{KleeorinMoss::2000, Hubbard:Brandenburg:2012}.

In the Babcock-Leighton framework, the poloidal field generation
results from the systematic tilt angle of bipolar magnetic regions.
While there is observational evidence that the average tilt angle
decreases with increasing cycle strength \citep{Dasi-Espuig:etal:2010,
  McClintockNorton::2013, Jiao:etal:2021}, catastrophic quenching is
not expected in this case \citep{KitchatinovOlemskoy::2011}.
Sufficiently strong magnetic field may also reduces the turbulent
magnetic diffusivity \citep[e.g.,][]{KleeorinRogachevskii::2007,
  GuerreroDikpati::2009}, which also affects the Babcock-Leighton
dynamo.

Magnetically induced changes of the large-scale differential rotation
\citep[zonal flows, cf.][]{LabonteHoward::1982} and of the meridional
circulation \citep[systematic inflows towards active regions,
  cf.][]{GizonDuvall::2001} have been observed \citep[see
  also][]{Hathaway:etal:2022}.  The zonal flows are probably too weak
to have a substantial effect on the dynamo mechanism.  The inflows
towards active regions explain part of the cyclic variation of the
meridional flow \citep{CameronSchussler::2012}. Their effects have
been studied in the framework of surface flux transport and
Babcock-Leighton models \citep{Martin-BeldaCameron::2017,
  NagyLemerle::2020}.  It was found that, in principle, the resulting
changes to the meridional flow can provide nonlinear saturation of the
dynamo.

Another nonlinearity which has been studied in the Babcock-Leighton
framework is 'latitudinal quenching' \citep{Jiang:2020,
  Talafha:etal:2022}. This effect is related to the observation that
the average latitudes of sunspots are located more poleward in strong
cycles as compared to weak cycles \citep{Waldmeier:1955,
  Solanki:etal:2008, Hathaway:2015}. Since bipolar regions at higher
latitudes contribute less to the flux crossing the equator and thus to
the buildup of the poloidal field \citep{JiangCameron::2014}, this
effect provides an amplitude-limiting nonlinearity.

\begin{figure}
  \centering
  \includegraphics[width=80mm]{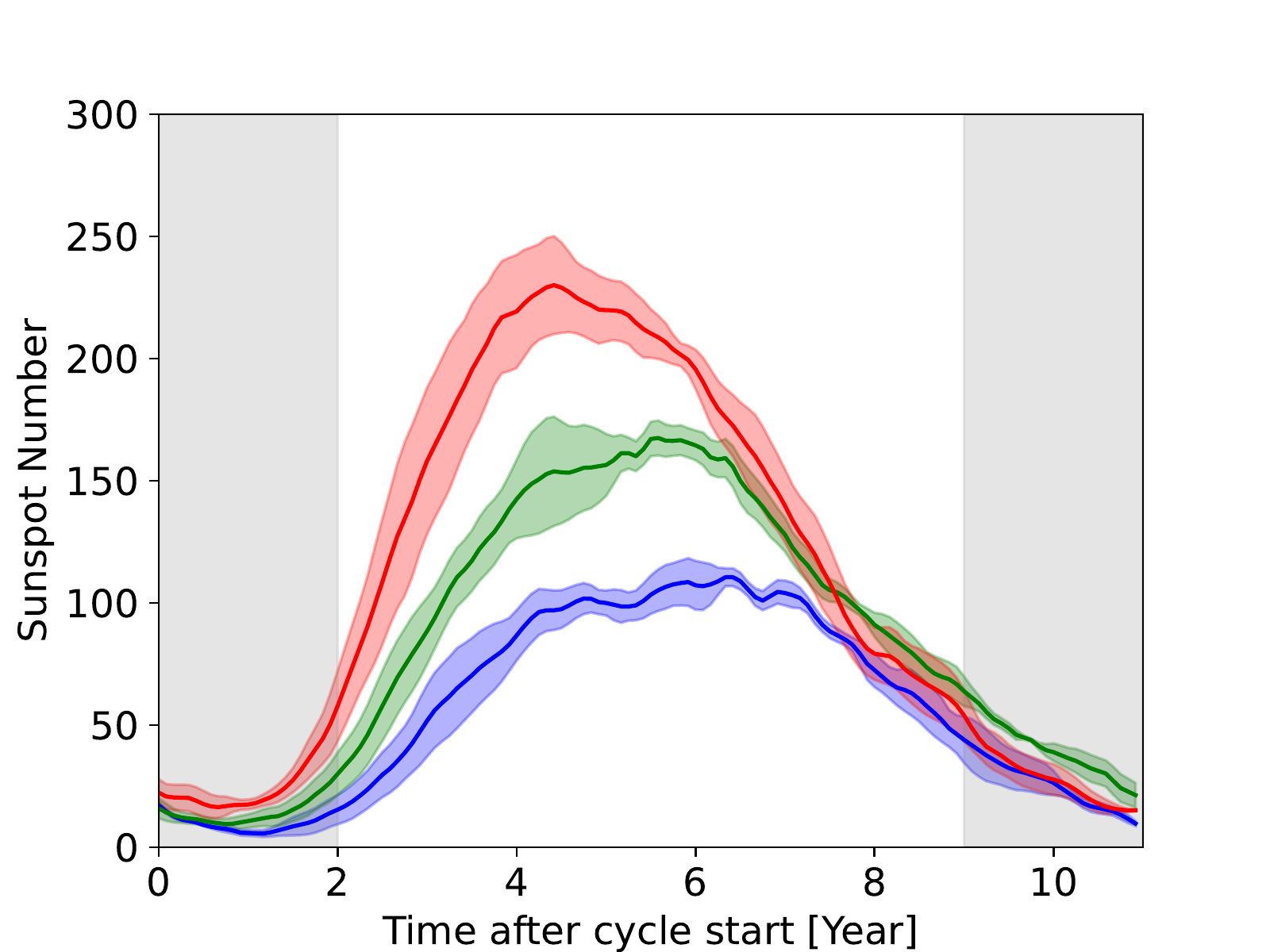}  
  \includegraphics[width=80mm]{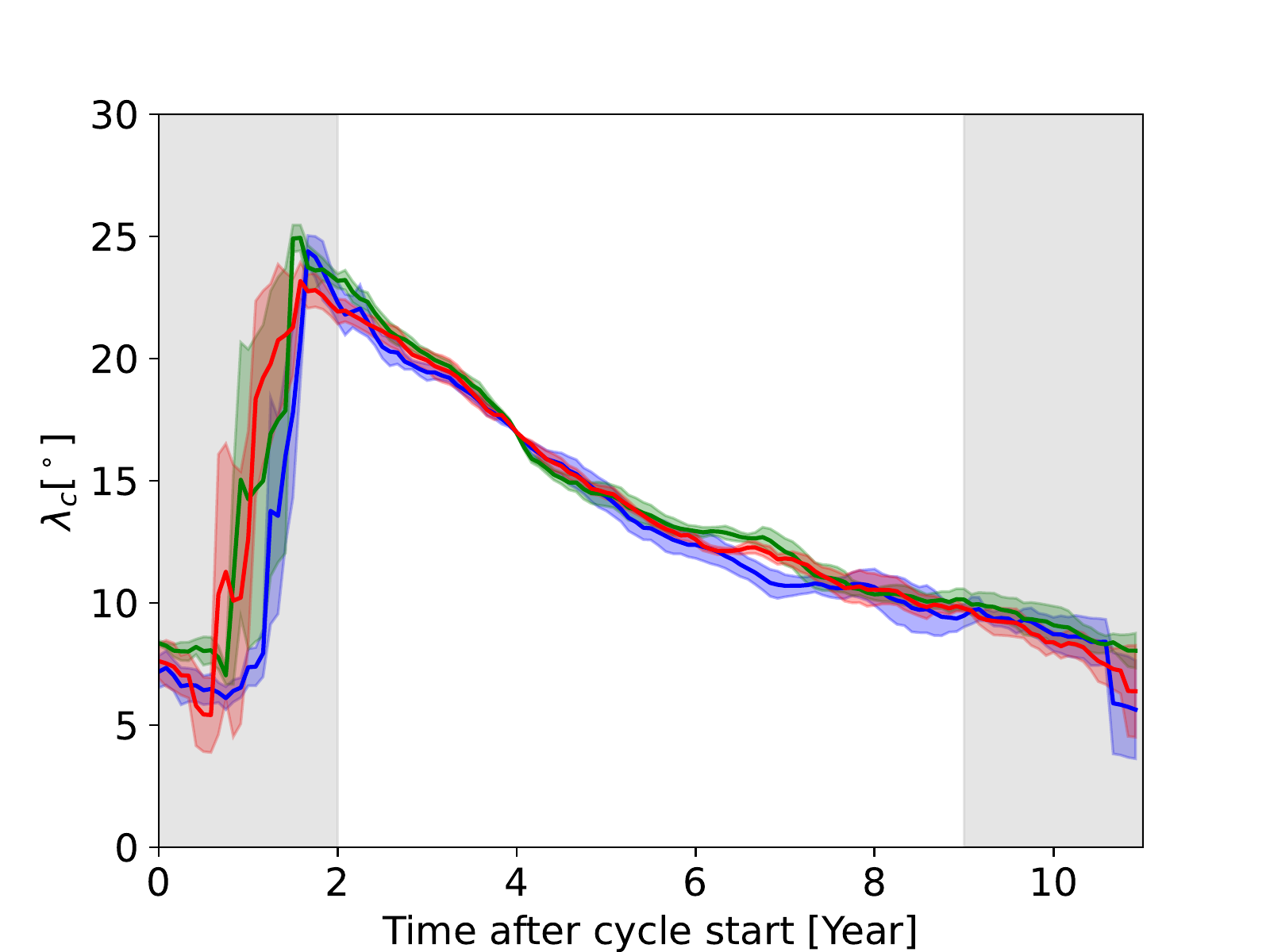}
  \includegraphics[width=80mm]{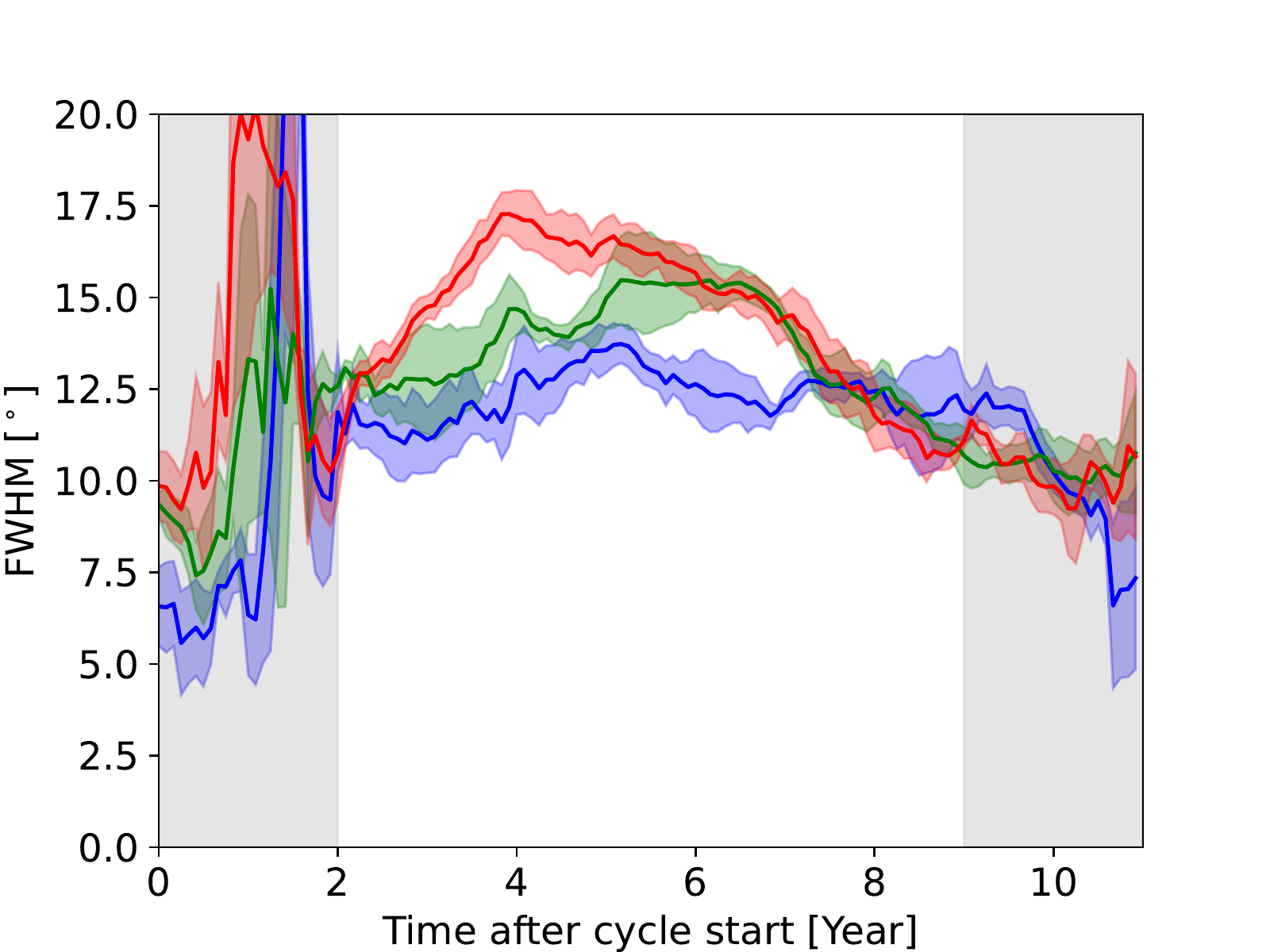}
  \caption{Properties of the sunspot zones as function of time from
    cycle start, based upon the historical sunspot record. Shown are
    the sunspot number (top panel), central latitude (middle panel),
    and full width at half maximum (lower panel) of the sunspot
    zones. The quantities are averaged over weak cycles ($n=12, 14,
    24, 16$, blue lines), intermediate cycles ($n= 15, 20, 17, 23$,
    green lines), and strong cycles ($n=18, 21, 22, 19$, red lines).
    Shading indicates the range covered by the $\pm 1$ stderr,
    calculated using the scatter between the four cycles in each
    group. }
\label{fig:Wald55}
\end{figure}

A closer look at the properties of flux emergence depending on cycle
strength yields further insight into the nature of latitude quenching.
In Fig.~\ref{fig:Wald55}, we used 13-month smoothed sunspot sunspot
numbers (Version 2) between solar cycles 12 and 24 from the SILSO data
base (https://www.sidc.be/silso/datafiles) and 12-month averages of
observed times, latitudes, and areas of sunspots given in the Royal
Greenwich Observatory and USAF/NOAA data bases (downloaded from
http://solarcyclescience.com/activeregions.html) to plot the sunspot
number, the central latitude, and the full width at half maximum
(FWHM) of the sunspot zones (``butterfly wings''), respectively, as
functions of time since the start of a cycle. Depending on their
peak sunspot number, four cycles each were put into groups of strong,
medium, and weak cycles.  Central latitude and FWHM of the sunspot
zones were determined from Gaussian fits of the sunspot data with
unsigned latitudes, thus merging both hemispheres.

Since the solar cycle is not perfectly periodic, comparing the
evolution of different cycles requires a definition of a reference
time for each cycle. One possibility is to fit the shape of the time
evolution of the sunspot number to a given functional
relationship. Using such a procedure, \citet{Hathaway::2011} found
that the central latitude of the sunspot belts propagates equatorward
in the same way for all cycles \citep[see also][]{Waldmeier:1939}. We
thus take the reference time as the instant at which the central
latitude is at $19^\circ$ and define the start of the cycle to be 4
years before that instant. These particular choices have no
significant impact on the analysis and the results.

The figure confirms earlier results of \citet{Waldmeier:1955}: (i) the
activity of stronger cycles rises faster and peaks earlier than that
of weaker cycles (often referred to as ``Waldmeier effect'') while the
declining phase is independent of cycle strength, and (ii) the time
profile of the propagation of the sunspot zones towards the equator is
independent of cycle strength \citep[see also][]{Hathaway::2011}.
Furthermore, the full wdth at half maximum of the sunspot zones in the
declining phase is also independent of cycle strength
\citep{Cameron:Schuessler:2016}, while stronger cycles show broader
sunspot zones \citep[see also][]{Mandal:etal:2017,
  BiswasKarak::2022}. These properties reveal three aspects of
latitude quenching, namely
\begin{enumerate}
\item{The Waldmeier effect has the consequence that flux emergence
  around cycle maxima on average occurs in higher latitudes for
  stronger cycles than for weaker cycles, thus being less effective
  for the buildup of the polar field. This corresponds to a negative
  feedback.}
\item{The broader wings of the sunspot zones during the maximum phases
  of stronger cycles have the opposite effect since more flux emerges
  in lower latitudes.}
\item{The fact that all three properties, sunspot number as well as
  central latitude and width of the sunspot zones, behave
  independently of cycle strength in the declining phase of the cycles
  means that, during this most critical phase for the buildup of the
  poloidal field, the amount of magnetic flux transfer transferred across
  the equator is independent of cycle strength, thus corresponding to
  negative feedback.}
\end{enumerate}
These three aspects reveal the action of an underlying nonlinearity
connected to flux emergence. It is plausible that the cycle strength
reflects the amount of toroidal magnetic flux in the convection zone
which is available for flux emergence. In strong cycles, latitudinal
differential rotation, which is steepest in mid latitudes, acts upon a
stronger poloidal field and thus produces more and stronger toroidal
magnetic flux. A nonlinearity matching the cycle properties discussed
above could be that flux emergence occurs when a critical field
strength of the order of the equipartition field strength is
exceeded. While the latitude drift of the sunspot zones is independent
of cycle strength (possibly being determined by a deep meridional flow
towards the equator, which is largely unaffected by the magnetic
field), the critical field strength is reached earlier in strong
cycles, meaning that more flux emerges at higher latitudes (point 1 of
the list above). At the same time, the critical field strength is
exceeded in a broader range of latitudes (point 2).  Consequently,
stronger cycles have lost a bigger part of their available toroidal
flux earlier in the cycle compared to weaker cycles, so that in the
later phases flux emergence and width of the sunspot zones become
independent of cycle strength.  \citet{BiswasKarak::2022} confirmed
this conjecture using a Babcock-Leighton flux-transport dynamo
model. They suggested that, in the declining phase of all cycles, flux
emergence compensates the increase of the toroidal field strength due
to the pileup of flux near the stagnation point of the equatorward
meridional flow. Consequently, the mean field strength remains near to
the critical field strength for flux emergence and all cycles decline
in the same way.

All mechanisms discussed in this section are plausible explanations
for the nonlinearity limiting the amplitude of the solar dynamo.
Distinguishing which combination of them acts on the Sun is an open
observational challenge.

\subsection{Long-term variability}\label{subsec:longterm}

\begin{figure}
  \includegraphics[width=\textwidth]{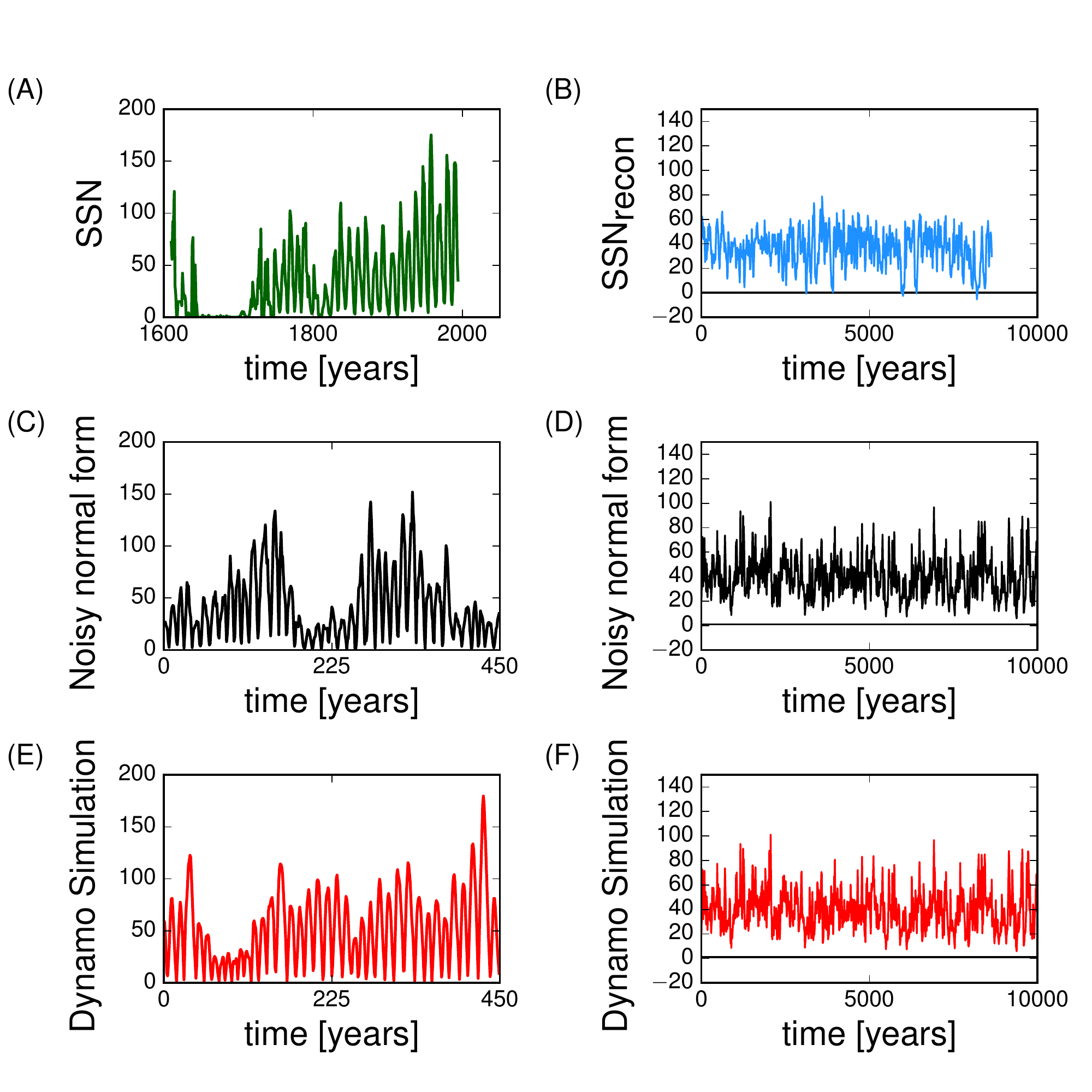}
  \caption{Observed and simulated records of solar activity. The
    historical sunspot record (sunspot group number, panel A) and the
    level of solar activity reconstructed from cosmogenic isotopes
    samples \citep{UsoskinGallet::2016} are shown in panels A and B,
    respectively. Panels C and D in the middle show random
    realizations of a normal-form model (near a Hopf bifurcation)
    model with noise, covering the same lengths of time as panels A and
    B. Panels E and F at the bottom shows realizations from a
    simple Babcock-Leighton dynamo model.  Figure taken from
    \citet{CameronSchuessler::2017a}.  }
\label{fig:LTV1}
\end{figure}

The sunspot cycle shows variability on a wide range of timescales
\citep[see reviews by][]{Usoskin:2017,Biswas:etal:2023}. Panel A
of Fig.~\ref{fig:LTV1} shows the historical sunspot record from
telescopic observations since the beginning of the 17th century. It
reveals significant cycle-to-cycle fluctuations of cycle strength
together with longer-term variability. Particularly conspicuous is the
period of very low sunspot activity between 1645 to 1715 and the
period of high average sunspot activity between about 1940 and
2006. Other such ``grand minima'' and ``grand maxima'' are found in
reconstructions of solar activity during the past millenia (albeit at
a coarser time resolution) on the basis of various records of
cosmogenic isotopes
\citep[e.g.,][]{SolankiUsoskin::2004,UsoskinGallet::2016}.  An example
of such a reconstruction is shown in panel B of Fig.~\ref{fig:LTV1}.

Observational studies of the rotational evolution (gyrochronology) of
solar-type stars \citep[e.g.,][]{MetcalfeEgeland::2016,
  vanSadersCeillier::2016, David:etal:2022, Metcalfe:etal:2022,
  Metcalfe:etal:2023} indicate that magnetic braking by a stellar wind
is significantly reduced for stars near or beyond the solar age, which
suggests a decline of their large-scale magnetic fields \citep[for a
  different view, see][]{Kotorashvili:etal:2023}.  This indicates
that, at its current rotation rate, the Sun could be  approaching a
transition point where its gobal dynamo switches off, so that at
present the excitation of the solar dynamo is only weakly
supercritical \citep[or even subcritical, as suggested
  by][]{Tripathi:etal:2021}.

The observationally well-studied rotation-activity relation for
magnetically active stars \citep[e.g.,][]{Brun:Browning:2017}
suggests the rotation rate as the relevant control parameter for
dynamo excitation.  For most dynamo models, the transition from
decaying field to excited oscillatory dynamo action corresponds to a
supercritical Hopf bifurcation \citep[e.g.,][]{TobiasWeiss::1995},
when a fixed point (equilibrium) spawns a limit cycle (oscillatory
solution). The behaviour of a a weakly excited nonlinear system near a
Hopf bifurcation is fully described by a generic normal-form model
that is independent of the specific properties of the system,
including also the nature of its nonlinearity
\citep[e.g.,][]{Guckenheimer:Holmes:1983}.

\cite{CameronSchuessler::2017a} applied this concept to the solar
dynamo and showed that the observed power spectrum of the solar cycle
(from the sunspot record and from the reconstruction based on
cosmogenic isotopes) is consistent with a noisy normal-form model
whose parameters are completely determined by observations.  The noise
in the system results from the large scatter in the observed tilt
angles of active regions, which is possibly associated with the
interaction of magnetic flux and convective motions
\citep{LongcopeFisher::1996}. As discussed in
Sections~\ref{subsec:btor} and \ref{subsec:bpol}, the tilt angle
scatter leads to randomness in the amount of flux transported across
the equator, and hence to randomness in the amount of toroidal field
generated for the subsequent cycle.

\begin{figure}
  \includegraphics[width=\textwidth]{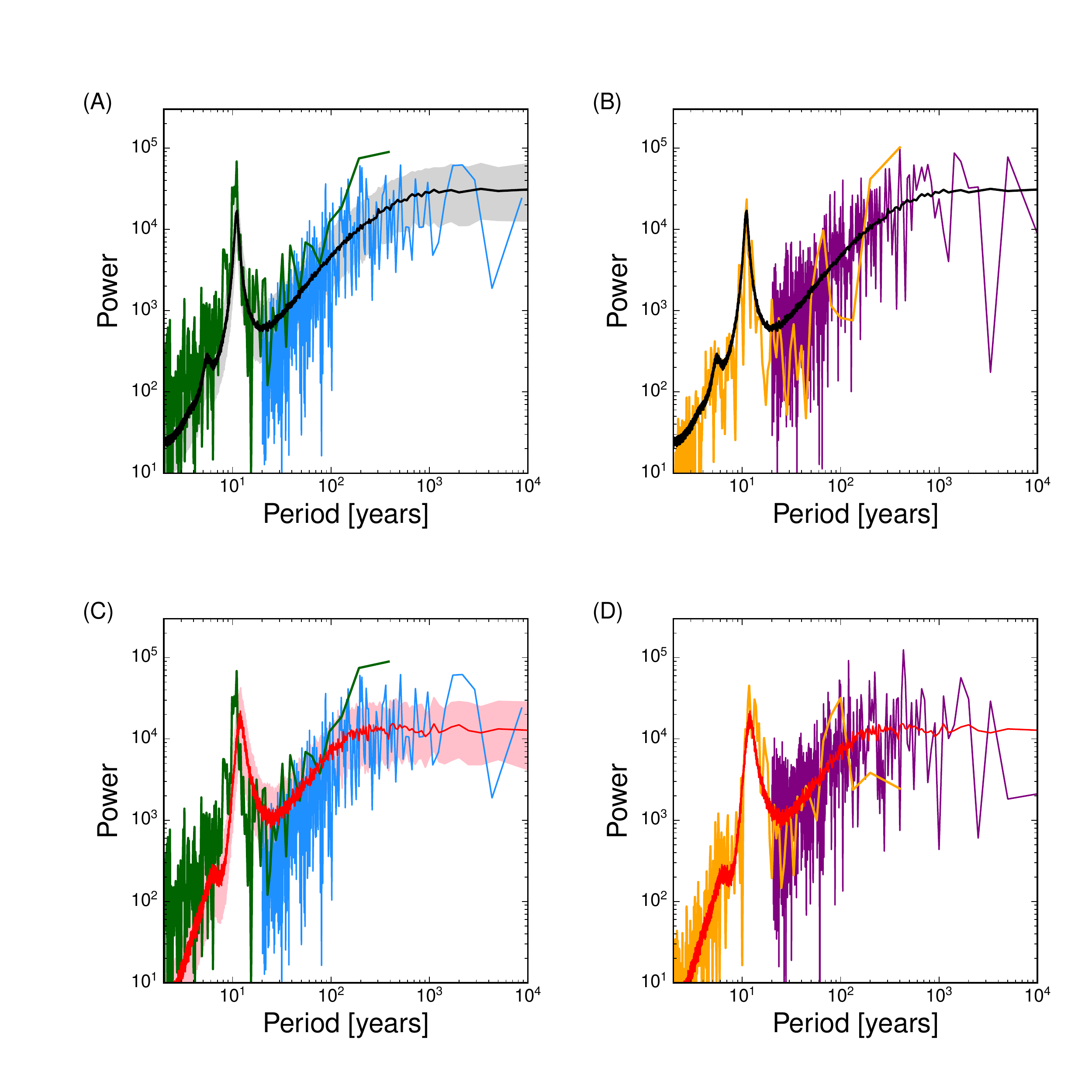}
  \caption{Power specta of empirical solar data and models. (A)
    Sunspot group number (green), and reconstruction from cosmogenic
    isotopes (blue), together with the expectation value from 10.000
    realizations of the normal form model (black curve) and the
    corresponding 25\% to 75\% quartiles (grey shading). (B) One
    realization of the normal form model.  (C) Same as (A), but
    compared to the expectation value and quartiles from realizations
    of an updated Babcock-Leighton model (black curve and pink
    shading). (D) One relization of the updated Babcock-Leighton
    model.  Figure taken from \cite{CameronSchuessler::2017a}. }
\label{fig:LTV2}
\end{figure}

Panels C and D of Fig.~\ref{fig:LTV1} show an example realization of
the normal-form model covering 10,000~years. Panel B of
Fig.~\ref{fig:LTV2} gives the corresponding temporal power spectrum,
which is consistent with the observed spectrum shown in panel A of
Fig.~\ref{fig:LTV2}. Moreover, the model also exhibits extended
periods of low activity (grand minima) whose statistical properties in
terms of the distributions of their lengths and the waiting times
between grand minima is consistent with the reconstructed record of
solar activity by \citet{UsoskinGallet::2016}.

Likewise, the updated 1D Babcock-Leighton dynamo model of
\citet{CameronSchussler::2017b} with slightly supercritical excitation
and including noise in the source term for the poloidal field also
yields time series that are statistically similar to the empirical
series (see panels E and F of Fig.~\ref{fig:LTV1}). The corresponding
power spectra in panels C and D of Fig.~\ref{fig:LTV2}) show a good
match to the observed spectra (panel A of Fig.~\ref{fig:LTV2}).  A
similar approach was taken by \citet{Kitchatinov:Nepomnyashchikh:2017}
and \citet{Kitchatinov:etal:2018}, who studied a more comprehensive
dynamo model.

\citet{CameronSchussler::2019} carried out a detailed analysis of
power pectra obtained with the generic noisy normal-form model
(incorporating only the 11/22~year base period) in order to evaluate
the statistical significance of periodicities inferred from the record
of reconstructed solar activity on the basis of cosmogenic isotopes.
They showed that power spectra from realizations covering 10,000~years
of simulated time (matching the length of the reconstructed solar
data) typically exhibit spectral peaks at various periods that are
qualitatively similar to those found in the solar data. Such peaks,
which result from the stochastic noise in the dynamo excitation, can
reach significance levels of $3\sigma$. These results cast doubt on
the proposition that seemingly significant periodicities such as the
\~90-year Gleissberg and the \~210-year de Vries ``cycles'' are
intrinsic to the solar dynamo and not just statistical
fluctuations. In fact, the sharpness of the corresponding peaks in the
power spectrum indicates a random origin since spectral peaks
representing intrinsic dynamo periodicities tend to be broadened owing
to the damping inherent to the dynamo process.

Apart from stochastic forcing as discussed above, long-term
variability in the dynamo process can also arise from magnetic
feedback on the flow and from time delays in the dynamo process. A
comprehensive review of models for the long-term variability has been
provided by \citet{Karak:2023}.

\subsection{Notes on predictability}\label{subsec:predictability}

Based on sufficient understanding of the global solar dynamo, the
observed state of the solar magnetic field can be used to predict its
future evolution \citep[for comprehensive reviews,
  see][]{Petrovay:2020, Bhowmik:etal:2023}. In the Babcock-Leighton
model, the cycle-to-cycle variability is directly related to the
amount of magnetic flux that gets across the equator. At the end of a
cycle, all this flux ends up in the polar regions and the amplitude of
the polar field is strongly correlated with the strength of the
subsequent cycle \citep[e.g.,][]{Kumar:etal:2021}. The physical basis
for this correlation is explained in Sec.~\ref{subsec:btor}.
Predicting the strength of a cycle thus becomes a matter of predicting
the polar field strength at the end of a cycle, which is almost
equivalent to predicting how much flux is transported across the
equator (see Sec.~\ref{subsec:bpol}). This can be achieved by
performing surface-flux-transport simulations using the actual
observations of latitude, magnetic flux, and tilt angle of each
emerging active region. Such simulations rather accurately reproduce
the amount of flux in each hemisphere and the resulting axial dipole
moment as a predictor for the subsequent cycle
\citep{JiangCameron::2015, Yeates:etal:2023}. A prediction for the
next cycle during an ongoing cycle (i.e., before all the active
regions of the cycle have emerged) can be made by including all active
regions which have been observed and performing Monte-Carlo
simulations of the effect of the active regions which have not yet
emerged, using their average properties
\citep{CameronJiang::2016}. The result of this approach is the
prediction of the polar field at the end of the cycle with error
estimates.  These can then be converted into predictions for the
strength of the subsequent cycle.

In any case, all predictions are limited by the inherent randomness of
the dynamo process \citep{Jiang:etal:2018, Kitchatinov:etal:2018}.
For instance, individual ``rogue active regions'' can have a strong
effect and may potentially even shut down the the large-scale dynamo
and initiate a grand minimum episode \citep{NagyLemerle::2017}.

\section{Outlook}\label{sec:outlook}

Considering the enormous range of scales and the complexity of
physical processes governing the interaction of turbulent convection,
differential rotation, and magnetic field in the solar convection
zone, it is rather surprising that a comparatively simple approach
such as the Babcock-Leighton scenario seems to provide such a
successful description of the solar dynamo process. An important
factor here is that key ingredients of the model, such as the
properties of bipolar magnetic regions (latitude range, tilt, flux
distribution, etc.) and surface flows, can be obtained by
observations. To a large extent, such observations are not available
for other stars, so that basic input is missing for the application of
the model. Furthermore, the underlying processes leading to flux
emergence, i.e., the roles of convective flows, magnetic buoyancy, and
instabilities deep in the convection zone, are largely not
understood. In the absence of direct observational evidence, 3D MHD
simulations seem to be the only possibility here. Considering the
impressive progress that simulations have seen during the last decade,
there is hope that the complexity of these processes will be better
understood in the not-too-far future, providing an even better basis
for simplified models such as the Babcock-Leighton approach. Still, we
need to exercise caution and be aware of the severe limitations of all
our efforts to understand the solar dynamo as lucidly expounded by
\citet{Parker:2009} and \citet{Spruit:2011, Spruit:2012}.

\bmhead{Acknowledgments} Sunspot data were obtained from the World
Data Center SILSO, Royal Observatory of Belgium, Brussels. The data
for the $aa$-index were obtained from
\url{https://isgi.unistra.fr/indices_aa.php}

\section*{Ethics Declarations}
{\bf Competing interests}

\medskip
The authors declare they have no conflicts of interest.

\bibliography{c+s_4.bbl}

\begin{thebibliography}{187}
\providecommand{\natexlab}[1]{#1}
\providecommand{\url}[1]{{#1}}
\providecommand{\urlprefix}{URL }
\providecommand{\doi}[1]{\url{https://doi.org/#1}}
\providecommand{\eprint}[2][]{\url{#2}}
 \bibcommenthead

\bibitem[{{Babcock}(1961)}]{Babcock:1961}
{Babcock} HW (1961) {The Topology of the Sun's Magnetic Field and the 22-year
  Cycle}. \apj 133:572. \doi{https://doi.org/10.1086/147060}

\bibitem[{{Babcock} and {Babcock}(1955)}]{Babcock:Babcock:1955}
{Babcock} HW, {Babcock} HD (1955) {The Sun's Magnetic Field, 1952-1954}. \apj
  121:349. \doi{10.1086/145994}

\bibitem[{{Barnes} et~al(1980){Barnes}, {Tryon}, and
  {Sargent}}]{Barnes:etal:1980}
{Barnes} JA, {Tryon} PV, {Sargent} IH.~H. (1980) {Sunspot cycle simulation
  using random noise}. In: {Pepin} RO, {Eddy} JA, {Merrill} RB (eds) The
  Ancient Sun: Fossil Record in the Earth, Moon and Meteorites, pp 159--163

\bibitem[{{Basu} and {Antia}(2019)}]{Basu:Antia:2019}
{Basu} S, {Antia} HM (2019) {Changes in Solar Rotation over Two Solar Cycles}.
  \apj 883(1):93. \doi{10.3847/1538-4357/ab3b57}

\bibitem[{{Baumann} et~al(2004){Baumann}, {Schmitt}, {Sch{\"u}ssler}, and
  {Solanki}}]{Baumann:etal:2004}
{Baumann} I, {Schmitt} D, {Sch{\"u}ssler} M, et~al (2004) {Evolution of the
  large-scale magnetic field on the solar surface: A parameter study}. \aap
  426:1075--1091. \doi{10.1051/0004-6361:20048024}

\bibitem[{{Bekki} and {Cameron}(2022)}]{Bekki:Cameron:2022}
{Bekki} Y, {Cameron} RH (2022) {Three-dimensional non-kinematic simulation of
  post-emergence evolution of bipolar magnetic regions and Babcock-Leighton
  dynamo of the Sun}. arXiv e-prints arXiv:2209.08178.
  \doi{10.48550/arXiv.2209.08178}

\bibitem[{{Bhowmik} and {Nandy}(2018)}]{Bhowmik:Nandy:2018}
{Bhowmik} P, {Nandy} D (2018) {Prediction of the strength and timing of sunspot
  cycle 25 reveal decadal-scale space environmental conditions}. Nature
  Communications 9:5209. \doi{10.1038/s41467-018-07690-0}

\bibitem[{{Bhowmik} et~al(2023){Bhowmik}, {Jiang}, {Upton}, {Lemerle}, and
  {Nandy}}]{Bhowmik:etal:2023}
{Bhowmik} P, {Jiang} J, {Upton} L, et~al (2023) {Physical Models for Solar
  Cycle Predictions}. \ssr p submitted. \doi{10.48550/arXiv.2303.12648}

\bibitem[{{Bieber} and {Rust}(1995)}]{BieberRust::1995}
{Bieber} JW, {Rust} DM (1995) {The Escape of Magnetic Flux from the Sun}. \apj
  453:911. \doi{10.1086/176451}

\bibitem[{{Birch} et~al(2016){Birch}, {Schunker}, {Braun}, {Cameron}, {Gizon},
  {Lo ptien}, and {Rempel}}]{Birch:etal:2016}
{Birch} AC, {Schunker} H, {Braun} DC, et~al (2016) {A low upper limit on the
  subsurface rise speed of solar active regions}. Science Advances
  2(7):e1600,557--e1600,557. \doi{10.1126/sciadv.1600557}

\bibitem[{{Biswas} et~al(2022){Biswas}, {Karak}, and
  {Cameron}}]{BiswasKarak::2022}
{Biswas} A, {Karak} BB, {Cameron} R (2022) {Toroidal Flux Loss due to Flux
  Emergence Explains why Solar Cycles Rise Differently but Decay in a Similar
  Way}. \prl 129(24):241102. \doi{10.1103/PhysRevLett.129.241102}

\bibitem[{{Biswas} et~al(2023){Biswas}, {Karak}, {Usoskin}, and
  {Weisshaar}}]{Biswas:etal:2023}
{Biswas} A, {Karak} B, {Usoskin} I, et~al (2023) {Long-term modulation of solar
  cycles}. \ssr p in press. \doi{10.48550/arXiv.2302.14845}

\bibitem[{{Brandenburg}(2005)}]{Brandenburg:2005}
{Brandenburg} A (2005) {The Case for a Distributed Solar Dynamo Shaped by
  Near-Surface Shear}. \apj 625(1):539--547. \doi{10.1086/429584}

\bibitem[{{Brandenburg} et~al(2023){Brandenburg}, {Elstner}, {Masada}, and
  {Pipin}}]{Brandenburg:etal:2023}
{Brandenburg} A, {Elstner} D, {Masada} Y, et~al (2023) {Turbulent processes and
  mean-field dynamo}. arXiv e-prints arXiv:2303.12425.
  \doi{10.48550/arXiv.2303.12425}

\bibitem[{{Brown} et~al(1989){Brown}, {Christensen-Dalsgaard}, {Dziembowski},
  {Goode}, {Gough}, and {Morrow}}]{Brown:etal:1989}
{Brown} TM, {Christensen-Dalsgaard} J, {Dziembowski} WA, et~al (1989)
  {Inferring the Sun's Internal Angular Velocity from Observed p-Mode Frequency
  Splittings}. \apj 343:526. \doi{10.1086/167727}

\bibitem[{{Browning} and {etal}(2023)}]{Browning:etal:2023}
{Browning} M, {etal} (2023) {tbd}. arXiv e-prints \doi{tbd}

\bibitem[{{Brun} and {Browning}(2017)}]{Brun:Browning:2017}
{Brun} AS, {Browning} MK (2017) {Magnetism, dynamo action and the solar-stellar
  connection}. \lrsp 14:4. \doi{10.1007/s41116-017-0007-8}

\bibitem[{{Caligari} et~al(1995){Caligari}, {Moreno-Insertis}, and
  {Sch{\"u}ssler}}]{Caligari:etal:1995}
{Caligari} P, {Moreno-Insertis} F, {Sch{\"u}ssler} M (1995) {Emerging Flux
  Tubes in the Solar Convection Zone. I. Asymmetry, Tilt, and Emergence
  Latitude}. \apj 441:886. \doi{10.1086/175410}

\bibitem[{{Cameron} and {Sch{\"u}ssler}(2015)}]{Cameron:Schuessler:2015}
{Cameron} R, {Sch{\"u}ssler} M (2015) {The crucial role of surface magnetic
  fields for the solar dynamo}. Science 347(6228):1333--1335.
  \doi{10.1126/science.1261470}

\bibitem[{{Cameron} and {Sch{\"u}ssler}(2012)}]{CameronSchussler::2012}
{Cameron} RH, {Sch{\"u}ssler} M (2012) {Are the strengths of solar cycles
  determined by converging flows towards the activity belts?} \aap 548:A57.
  \doi{10.1051/0004-6361/201219914}

\bibitem[{{Cameron} and {Sch{\"u}ssler}(2016)}]{Cameron:Schuessler:2016}
{Cameron} RH, {Sch{\"u}ssler} M (2016) {The turbulent diffusion of toroidal
  magnetic flux as inferred from properties of the sunspot butterfly diagram}.
  \aap 591:A46. \doi{10.1051/0004-6361/201527284}

\bibitem[{{Cameron} and
  {Sch{\"u}ssler}(2017{\natexlab{a}})}]{CameronSchussler::2017b}
{Cameron} RH, {Sch{\"u}ssler} M (2017{\natexlab{a}}) {An update of Leighton's
  solar dynamo model}. \aap 599:A52.
  \doi{https://doi.org/10.1051/0004-6361/201629746}

\bibitem[{{Cameron} and
  {Sch{\"u}ssler}(2017{\natexlab{b}})}]{CameronSchuessler::2017a}
{Cameron} RH, {Sch{\"u}ssler} M (2017{\natexlab{b}}) {Understanding Solar Cycle
  Variability}. \apj 843(2):111. \doi{10.3847/1538-4357/aa767a}

\bibitem[{{Cameron} and {Sch{\"u}ssler}(2019)}]{CameronSchussler::2019}
{Cameron} RH, {Sch{\"u}ssler} M (2019) {Solar activity: periodicities beyond 11
  years are consistent with random forcing}. \aap 625:A28.
  \doi{https://doi.org/10.1051/0004-6361/201935290}

\bibitem[{{Cameron} and {Sch{\"u}ssler}(2020)}]{CameronSchuessler::2020}
{Cameron} RH, {Sch{\"u}ssler} M (2020) {Loss of toroidal magnetic flux by
  emergence of bipolar magnetic regions}. \aap 636:A7.
  \doi{10.1051/0004-6361/201937281}

\bibitem[{{Cameron} et~al(2012){Cameron}, {Schmitt}, {Jiang}, and
  {I{\c{s}}{\i}k}}]{Cameron:etal:2012}
{Cameron} RH, {Schmitt} D, {Jiang} J, et~al (2012) {Surface flux evolution
  constraints for flux transport dynamos}. \aap 542:A127.
  \doi{10.1051/0004-6361/201218906}

\bibitem[{{Cameron} et~al(2013){Cameron}, {Dasi-Espuig}, {Jiang},
  {I{\c{s}}{\i}k}, {Schmitt}, and {Sch{\"u}ssler}}]{CameronDasi-Espuig::2013}
{Cameron} RH, {Dasi-Espuig} M, {Jiang} J, et~al (2013) {Limits to solar cycle
  predictability: Cross-equatorial flux plumes}. \aap 557:A141.
  \doi{https://doi.org/10.1051/0004-6361/201321981}

\bibitem[{{Cameron} et~al(2014){Cameron}, {Jiang}, {Sch{\"u}ssler}, and
  {Gizon}}]{CameronJiang::2014}
{Cameron} RH, {Jiang} J, {Sch{\"u}ssler} M, et~al (2014) {Physical causes of
  solar cycle amplitude variability}. \jgr 119(2):680--688.
  \doi{https://doi.org/10.1002/2013JA019498}

\bibitem[{{Cameron} et~al(2016){Cameron}, {Jiang}, and
  {Sch{\"u}ssler}}]{CameronJiang::2016}
{Cameron} RH, {Jiang} J, {Sch{\"u}ssler} M (2016) {Solar Cycle 25: Another
  Moderate Cycle?} \apjl 823(2):L22. \doi{10.3847/2041-8205/823/2/L22}

\bibitem[{{Cameron} et~al(2018){Cameron}, {Duvall}, {Sch{\"u}ssler}, and
  {Schunker}}]{CameronDuvall::2018}
{Cameron} RH, {Duvall} TL, {Sch{\"u}ssler} M, et~al (2018) {Observing and
  modeling the poloidal and toroidal fields of the solar dynamo}. \aap 609:A56.
  \doi{https://doi.org/10.1051/0004-6361/201731481}

\bibitem[{{Cattaneo} and {Vainshtein}(1991)}]{CattaneoVainshtein::1991}
{Cattaneo} F, {Vainshtein} SI (1991) {Suppression of Turbulent Transport by a
  Weak Magnetic Field}. \apjl 376:L21. \doi{10.1086/186093}

\bibitem[{{Centeno}(2012)}]{Centeno:2012}
{Centeno} R (2012) {The Naked Emergence of Solar Active Regions Observed with
  SDO/HMI}. \apj 759(1):72. \doi{10.1088/0004-637X/759/1/72}

\bibitem[{{Charbonneau}(2020)}]{Charbonneau:2020}
{Charbonneau} P (2020) {Dynamo models of the solar cycle}. \lrsp 17(1):4.
  \doi{10.1007/s41116-020-00025-6}

\bibitem[{{Charbonneau} and {MacGregor}(1997)}]{Charbonneau:MacGregor:1997}
{Charbonneau} P, {MacGregor} KB (1997) {Solar Interface Dynamos. II. Linear,
  Kinematic Models in Spherical Geometry}. \apj 486(1):502--520.
  \doi{10.1086/304485}

\bibitem[{{Chatterjee} et~al(2004){Chatterjee}, {Nandy}, and
  {Choudhuri}}]{Chatterjee:etal:2004}
{Chatterjee} P, {Nandy} D, {Choudhuri} AR (2004) {Full-sphere simulations of a
  circulation-dominated solar dynamo: Exploring the parity issue}. \aap
  427:1019--1030. \doi{10.1051/0004-6361:20041199}

\bibitem[{{Chen} et~al(2017){Chen}, {Rempel}, and {Fan}}]{Chen:etal:2017}
{Chen} F, {Rempel} M, {Fan} Y (2017) {Emergence of Magnetic Flux Generated in a
  Solar Convective Dynamo. I. The Formation of Sunspots and Active Regions, and
  The Origin of Their Asymmetries}. \apj 846(2):149.
  \doi{10.3847/1538-4357/aa85a0}

\bibitem[{{Choudhuri} and {Gilman}(1987)}]{Choudhuri:Gilman:1987}
{Choudhuri} AR, {Gilman} PA (1987) {The Influence of the Coriolis Force on Flux
  Tubes Rising through the Solar Convection Zone}. \apj 316:788.
  \doi{10.1086/165243}

\bibitem[{{Choudhuri} and {Hazra}(2016)}]{Choudhuri:Hazra:2016}
{Choudhuri} AR, {Hazra} G (2016) {The treatment of magnetic buoyancy in flux
  transport dynamo models}. Advances in Space Research 58(8):1560--1570.
  \doi{10.1016/j.asr.2016.03.015}

\bibitem[{{Choudhuri} et~al(1995){Choudhuri}, {Sch{\"u}ssler}, and
  {Dikpati}}]{Choudhuri:etal:1995}
{Choudhuri} AR, {Sch{\"u}ssler} M, {Dikpati} M (1995) {The solar dynamo with
  meridional circulation.} \aap 303:L29

\bibitem[{{Christensen-Dalsgaard} et~al(2011){Christensen-Dalsgaard},
  {Monteiro}, {Rempel}, and {Thompson}}]{Christensen-Dalsgaard:etal:2011}
{Christensen-Dalsgaard} J, {Monteiro} MJPFG, {Rempel} M, et~al (2011) {A more
  realistic representation of overshoot at the base of the solar convective
  envelope as seen by helioseismology}. \mnras 414(2):1158--1174.
  \doi{10.1111/j.1365-2966.2011.18460.x}

\bibitem[{{Cowling}(1953)}]{Cowling:1953}
{Cowling} TG (1953) {Solar Electrodynamics}. In: {Kuiper} GP (ed) The Sun.
  University of Chicago Press, p 532

\bibitem[{{Dasi-Espuig} et~al(2010){Dasi-Espuig}, {Solanki}, {Krivova},
  {Cameron}, and {Pe{\~n}uela}}]{Dasi-Espuig:etal:2010}
{Dasi-Espuig} M, {Solanki} SK, {Krivova} NA, et~al (2010) {Sunspot group tilt
  angles and the strength of the solar cycle}. \aap 518:A7.
  \doi{10.1051/0004-6361/201014301}

\bibitem[{{David} et~al(2022){David}, {Angus}, {Curtis}, {van Saders},
  {Colman}, {Contardo}, {Lu}, and {Zinn}}]{David:etal:2022}
{David} TJ, {Angus} R, {Curtis} JL, et~al (2022) {Further Evidence of Modified
  Spin-down in Sun-like Stars: Pileups in the Temperature-Period Distribution}.
  \apj 933(1):114. \doi{10.3847/1538-4357/ac6dd3}

\bibitem[{{DeVore} et~al(1984){DeVore}, {Boris}, and
  {Sheeley}}]{DeVore:etal:1984}
{DeVore} CR, {Boris} JP, {Sheeley} JN.~R. (1984) {The concentration of the
  large-scale solar magnetic field by a meridional surface flow}. \solphys
  92(1-2):1--14. \doi{10.1007/BF00157230}

\bibitem[{{Dikpati} and {Charbonneau}(1999)}]{Dikpati:Charbonneau:1999}
{Dikpati} M, {Charbonneau} P (1999) {A Babcock-Leighton Flux Transport Dynamo
  with Solar-like Differential Rotation}. \apj 518(1):508--520.
  \doi{10.1086/307269}

\bibitem[{{Durney}(1995)}]{Durney:1995}
{Durney} BR (1995) {On a Babcock-Leighton dynamo model with a deep-seated
  generating layer for the toroidal magnetic field}. \solphys 160(2):213--235.
  \doi{10.1007/BF00732805}

\bibitem[{{Durney}(1997)}]{Durney:1997}
{Durney} BR (1997) {On a Babcock-Leighton Solar Dynamo Model with a Deep-seated
  Generating Layer for the Toroidal Magnetic Field. IV.} \apj
  486(2):1065--1077. \doi{10.1086/304546}

\bibitem[{{Durrant} et~al(2004){Durrant}, {Turner}, and
  {Wilson}}]{DurrantTurner::2004}
{Durrant} CJ, {Turner} JPR, {Wilson} PR (2004) {The Mechanism involved in the
  Reversals of the Sun's Polar Magnetic Fields}. \solphys 222(2):345--362.
  \doi{10.1023/B:SOLA.0000043577.33961.82}

\bibitem[{{Duvall}(1979)}]{Duvall:1979}
{Duvall} JT.~L. (1979) {Large-scale solar velocity fields.} \solphys
  63(1):3--15. \doi{10.1007/BF00155690}

\bibitem[{{Fan}(2021)}]{Fan:2021}
{Fan} Y (2021) {Magnetic fields in the solar convection zone}. \lrsp 18(1):5.
  \doi{10.1007/s41116-021-00031-2}

\bibitem[{{Fan} and {Fang}(2014)}]{Fan:Fang:2014}
{Fan} Y, {Fang} F (2014) {A Simulation of Convective Dynamo in the Solar
  Convective Envelope: Maintenance of the Solar-like Differential Rotation and
  Emerging Flux}. \apj 789(1):35. \doi{10.1088/0004-637X/789/1/35}

\bibitem[{{Fan} et~al(1994){Fan}, {Fisher}, and {McClymont}}]{Fan:etal:1994}
{Fan} Y, {Fisher} GH, {McClymont} AN (1994) {Dynamics of Emerging Active Region
  Flux Loops}. \apj 436:907. \doi{10.1086/174967}

\bibitem[{{Ferriz-Mas} and {Sch{\"u}ssler}(1993)}]{Ferriz-Mas:Schuessler:1993}
{Ferriz-Mas} A, {Sch{\"u}ssler} M (1993) {Instabilities of magnetic flux tubes
  in a stellar convection zone I. Equatorial flux rings in differentially
  rotating stars}. Geophysical and Astrophysical Fluid Dynamics 72(1):209--247.
  \doi{10.1080/03091929308203613}

\bibitem[{{Ferriz-Mas} and {Sch{\"u}ssler}(1995)}]{Ferriz-Mas:Schuessler:1995}
{Ferriz-Mas} A, {Sch{\"u}ssler} M (1995) {Instabilities of magnetic flux tubes
  in a stellar convection zone II. Flux rings outside the equatorial plane}.
  Geophysical and Astrophysical Fluid Dynamics 81(3):233--265

\bibitem[{{Gaizauskas} et~al(1983){Gaizauskas}, {Harvey}, {Harvey}, and
  {Zwaan}}]{Gaizauskas:etal:1983}
{Gaizauskas} V, {Harvey} KL, {Harvey} JW, et~al (1983) {Large-scale patterns
  formed by solar active regions during the ascending phase of cycle 21}. \apj
  265:1056--1065. \doi{10.1086/160747}

\bibitem[{{Galloway} and {Weiss}(1981)}]{Galloway:Weiss:1981}
{Galloway} DJ, {Weiss} NO (1981) {Convection and magnetic fields in stars}.
  \apj 243:945--953. \doi{10.1086/158659}

\bibitem[{{Giovanelli}(1985)}]{Giovanelli:1985}
{Giovanelli} RG (1985) {The sunspot cycle and solar magnetic fields. I - The
  mechanism as inferred from observation. II - The interaction of flux tubes
  with the convection zone}. Australian Journal of Physics 38:1045--1089.
  \doi{10.1071/PH851045}

\bibitem[{{Gizon} et~al(2001){Gizon}, {Duvall}, and
  {Larsen}}]{GizonDuvall::2001}
{Gizon} L, {Duvall} JT.~L., {Larsen} RM (2001) {Probing Surface Flows and
  Magnetic Activity with Time-Distance Helioseismology}. In: {Brekke} P,
  {Fleck} B, {Gurman} JB (eds) Recent Insights into the Physics of the Sun and
  Heliosphere: Highlights from SOHO and Other Space Missions, p 189

\bibitem[{{Gizon} et~al(2020){Gizon}, {Cameron}, {Pourabdian}, {Liang},
  {Fournier}, {Birch}, and {Hanson}}]{GizonCameron::2020}
{Gizon} L, {Cameron} R, {Pourabdian} M, et~al (2020) {Meridional flow in the
  Sun{\textquoteright}s convection zone is a single cell in each hemisphere}.
  Science 368(6498):1469--1472. \doi{https://doi.org/10.1126/science.aaz7119}

\bibitem[{{Gottschling} et~al(2021){Gottschling}, {Schunker}, {Birch},
  {L{\"o}ptien}, and {Gizon}}]{Gottschling:etal:2021}
{Gottschling} N, {Schunker} H, {Birch} AC, et~al (2021) {Evolution of solar
  surface inflows around emerging active regions}. \aap 652:A148.
  \doi{10.1051/0004-6361/202140324}

\bibitem[{Guckenheimer and Holmes(1983)}]{Guckenheimer:Holmes:1983}
Guckenheimer J, Holmes P (1983) Nonlinear oscillations, dynamical systems, and
  bifurcations of vector fields. Applied mathematical sciences, Springer, New
  York

\bibitem[{{Guerrero} and {de Gouveia Dal Pino}(2007)}]{Guerrero:Gouveia:2007}
{Guerrero} G, {de Gouveia Dal Pino} EM (2007) {How does the shape and thickness
  of the tachocline affect the distribution of the toroidal magnetic fields in
  the solar dynamo?} \aap 464(1):341--349. \doi{10.1051/0004-6361:20065834}

\bibitem[{{Guerrero} et~al(2009){Guerrero}, {Dikpati}, and {de Gouveia Dal
  Pino}}]{GuerreroDikpati::2009}
{Guerrero} G, {Dikpati} M, {de Gouveia Dal Pino} EM (2009) {The Role of
  Diffusivity Quenching in Flux-transport Dynamo Models}. \apj 701(1):725--736.
  \doi{10.1088/0004-637X/701/1/725}

\bibitem[{{Hale} et~al(1919){Hale}, {Ellerman}, {Nicholson}, and
  {Joy}}]{HaleEllerman::1919}
{Hale} GE, {Ellerman} F, {Nicholson} SB, et~al (1919) {The Magnetic Polarity of
  Sun-Spots}. \apj 49:153. \doi{10.1086/142452}

\bibitem[{{Hanasoge}(2022)}]{Hanasoge:2022}
{Hanasoge} SM (2022) {Surface and interior meridional circulation in the Sun}.
  \lrsp 19(1):3. \doi{10.1007/s41116-022-00034-7}

\bibitem[{{Harvey}(1993)}]{Harvey::1993}
{Harvey} KL (1993) {Magnetic bipoles on the Sun}. PhD thesis, University of
  Utrecht, Netherlands

\bibitem[{{Harvey} et~al(1975){Harvey}, {Harvey}, and
  {Martin}}]{HarveyHarvey::1975}
{Harvey} KL, {Harvey} JW, {Martin} SF (1975) {Ephemeral Active Regions in 1970
  and 1973}. \solphys 40(1):87--102. \doi{10.1007/BF00183154}

\bibitem[{{Hathaway}(2011)}]{Hathaway::2011}
{Hathaway} DH (2011) {A Standard Law for the Equatorward Drift of the Sunspot
  Zones}. \solphys 273(1):221--230. \doi{10.1007/s11207-011-9837-z}

\bibitem[{{Hathaway}(2015)}]{Hathaway:2015}
{Hathaway} DH (2015) {The Solar Cycle}. \lrsp 12(1):4.
  \doi{10.1007/lrsp-2015-4}

\bibitem[{{Hathaway} and {Upton}(2016)}]{Hathaway:Upton:2016}
{Hathaway} DH, {Upton} LA (2016) {Predicting the amplitude and hemispheric
  asymmetry of solar cycle 25 with surface flux transport}. Journal of
  Geophysical Research (Space Physics) 121(11):10,744--10,753.
  \doi{10.1002/2016JA023190}

\bibitem[{{Hathaway} et~al(2022){Hathaway}, {Upton}, and
  {Mahajan}}]{Hathaway:etal:2022}
{Hathaway} DH, {Upton} LA, {Mahajan} SS (2022) {Variations in differential
  rotation and meridional flow within the Sun's surface shear layer
  1996{\textendash}2022}. Frontiers in Astronomy and Space Sciences 9:419.
  \doi{10.3389/fspas.2022.1007290}

\bibitem[{{Hazra}(2021)}]{Hazra:2021}
{Hazra} G (2021) {Recent advances in the 3D kinematic Babcock-Leighton solar
  dynamo modeling}. Journal of Astrophysics and Astronomy 42(2):22.
  \doi{10.1007/s12036-021-09738-y}

\bibitem[{{Hazra} and {Miesch}(2018)}]{Hazra:Miesch:2018}
{Hazra} G, {Miesch} MS (2018) {Incorporating Surface Convection into a 3D
  Babcock-Leighton Solar Dynamo Model}. \apj 864(2):110.
  \doi{10.3847/1538-4357/aad556}

\bibitem[{{Hazra} et~al(2017){Hazra}, {Choudhuri}, and
  {Miesch}}]{Hazra:etal:2017}
{Hazra} G, {Choudhuri} AR, {Miesch} MS (2017) {A Theoretical Study of the
  Build-up of the Sun{\textquoteright}s Polar Magnetic Field by using a 3D
  Kinematic Dynamo Model}. \apj 835(1):39. \doi{10.3847/1538-4357/835/1/39}

\bibitem[{{Hazra} et~al(2023){Hazra}, {Nandy}, {Kitchatinov}, and
  {Choudhuri}}]{Hazra:etal:2023}
{Hazra} G, {Nandy} D, {Kitchatinov} L, et~al (2023) {Mean field models of flux
  transport dynamo and meridional circulation in the Sun and stars}. \ssr p
  submitted. \doi{10.48550/arXiv.2302.09390}

\bibitem[{{Hotta}(2017)}]{Hotta:2017}
{Hotta} H (2017) {Solar Overshoot Region and Small-scale Dynamo with Realistic
  Energy Flux}. \apj 843(1):52. \doi{10.3847/1538-4357/aa784b}

\bibitem[{{Howard}(1979)}]{Howard:1979}
{Howard} R (1979) {Evidence for large-scale velocity features on the sun.}
  \apjl 228:L45--L50. \doi{10.1086/182900}

\bibitem[{{Howard}(1996)}]{Howard:1996}
{Howard} RF (1996) {Solar Active Regions As Diagnostics of Subsurface
  Conditions}. \araa 34:75--110. \doi{10.1146/annurev.astro.34.1.75}

\bibitem[{{Howe}(2009)}]{Howe:2009}
{Howe} R (2009) {Solar Interior Rotation and its Variation}. \lrsp 6(1):1.
  \doi{10.12942/lrsp-2009-1}

\bibitem[{{Hubbard} and {Brandenburg}(2012)}]{Hubbard:Brandenburg:2012}
{Hubbard} A, {Brandenburg} A (2012) {Catastrophic Quenching in
  {\ensuremath{\alpha}}{\ensuremath{\Omega}} Dynamos Revisited}. \apj
  748(1):51. \doi{10.1088/0004-637X/748/1/51}

\bibitem[{{Isik} and {etal}(2023)}]{Isik:etal:2023}
{Isik} E, {etal} (2023) {tbd}. arXiv e-prints \doi{tbd}

\bibitem[{{Jeffers} et~al(2022){Jeffers}, {Cameron}, {Marsden}, {Boro Saikia},
  {Folsom}, {Jardine}, {Morin}, {Petit}, {See}, {Vidotto}, {Wolter}, and
  {Mittag}}]{JeffersCameron::2022}
{Jeffers} SV, {Cameron} RH, {Marsden} SC, et~al (2022) {The crucial role of
  surface magnetic fields for stellar dynamos: $\epsilon$ Eridani, 61 Cygni A,
  and the Sun}. \aap 661:A152.
  \doi{https://doi.org/10.1051/0004-6361/202142202}

\bibitem[{{Jiang}(2020)}]{Jiang:2020}
{Jiang} J (2020) {Nonlinear Mechanisms that Regulate the Solar Cycle
  Amplitude}. \apj 900(1):19. \doi{10.3847/1538-4357/abaa4b}

\bibitem[{{Jiang} et~al(2014{\natexlab{a}}){Jiang}, {Cameron}, and
  {Sch{\"u}ssler}}]{JiangCameron::2014}
{Jiang} J, {Cameron} RH, {Sch{\"u}ssler} M (2014{\natexlab{a}}) {Effects of the
  Scatter in Sunspot Group Tilt Angles on the Large-scale Magnetic Field at the
  Solar Surface}. \apj 791(1):5. \doi{10.1088/0004-637X/791/1/5}

\bibitem[{{Jiang} et~al(2014{\natexlab{b}}){Jiang}, {Hathaway}, {Cameron},
  {Solanki}, {Gizon}, and {Upton}}]{Jiang:etal:2014}
{Jiang} J, {Hathaway} DH, {Cameron} RH, et~al (2014{\natexlab{b}}) {Magnetic
  Flux Transport at the Solar Surface}. \ssr 186(1-4):491--523.
  \doi{10.1007/s11214-014-0083-1}

\bibitem[{{Jiang} et~al(2015){Jiang}, {Cameron}, and
  {Sch{\"u}ssler}}]{JiangCameron::2015}
{Jiang} J, {Cameron} RH, {Sch{\"u}ssler} M (2015) {The Cause of the Weak Solar
  Cycle 24}. \apjl 808(1):L28.
  \doi{https://doi.org/10.1088/2041-8205/808/1/L28}

\bibitem[{{Jiang} et~al(2018){Jiang}, {Wang}, {Jiao}, and
  {Cao}}]{Jiang:etal:2018}
{Jiang} J, {Wang} JX, {Jiao} QR, et~al (2018) {Predictability of the Solar
  Cycle Over One Cycle}. \apj 863(2):159. \doi{10.3847/1538-4357/aad197}

\bibitem[{{Jiao} et~al(2021){Jiao}, {Jiang}, and {Wang}}]{Jiao:etal:2021}
{Jiao} Q, {Jiang} J, {Wang} ZF (2021) {Sunspot tilt angles revisited:
  Dependence on the solar cycle strength}. \aap 653:A27.
  \doi{10.1051/0004-6361/202141215}

\bibitem[{{Karak}(2023)}]{Karak:2023}
{Karak} B (2023) {Models for the long-term variations of solar activity}. \lrsp
  p in press

\bibitem[{{Karak} and {Cameron}(2016)}]{Karak:Cameron:2016}
{Karak} BB, {Cameron} R (2016) {Babcock-Leighton Solar Dynamo: The Role of
  Downward Pumping and the Equatorward Propagation of Activity}. \apj
  832(1):94. \doi{10.3847/0004-637X/832/1/94}

\bibitem[{{Karak} and {Miesch}(2017)}]{Karak:Miesch:2017}
{Karak} BB, {Miesch} M (2017) {Solar Cycle Variability Induced by Tilt Angle
  Scatter in a Babcock-Leighton Solar Dynamo Model}. \apj 847(1):69.
  \doi{10.3847/1538-4357/aa8636}

\bibitem[{{Kitchatinov} and
  {Nepomnyashchikh}(2017)}]{Kitchatinov:Nepomnyashchikh:2017}
{Kitchatinov} L, {Nepomnyashchikh} A (2017) {How supercritical are stellar
  dynamos, or why do old main-sequence dwarfs not obey gyrochronology?} \mnras
  470(3):3124--3130. \doi{10.1093/mnras/stx1473}

\bibitem[{{Kitchatinov} and
  {Olemskoy}(2011{\natexlab{a}})}]{KitchatinovOlemskoy::2011}
{Kitchatinov} LL, {Olemskoy} SV (2011{\natexlab{a}}) {Alleviation of
  catastrophic quenching in solar dynamo model with nonlocal alpha-effect}. \an
  332(5):496. \doi{10.1002/asna.201011549}

\bibitem[{{Kitchatinov} and
  {Olemskoy}(2011{\natexlab{b}})}]{Kitchatinov:Olemskoy:2011}
{Kitchatinov} LL, {Olemskoy} SV (2011{\natexlab{b}}) {Does the Babcock-Leighton
  mechanism operate on the Sun?} Astronomy Letters 37(9):656--658.
  \doi{10.1134/S0320010811080031}

\bibitem[{{Kitchatinov} et~al(2018){Kitchatinov}, {Mordvinov}, and
  {Nepomnyashchikh}}]{Kitchatinov:etal:2018}
{Kitchatinov} LL, {Mordvinov} AV, {Nepomnyashchikh} AA (2018) {Modelling
  variability of solar activity cycles}. \aap 615:A38.
  \doi{10.1051/0004-6361/201732549}

\bibitem[{{Kleeorin} and {Rogachevskii}(2007)}]{KleeorinRogachevskii::2007}
{Kleeorin} N, {Rogachevskii} I (2007) {Nonlinear turbulent magnetic diffusion
  and effective drift velocity of a large-scale magnetic field in
  two-dimensional magnetohydrodynamic turbulence}. \pre 75(6):066315.
  \doi{10.1103/PhysRevE.75.066315}

\bibitem[{{Kleeorin} et~al(2000){Kleeorin}, {Moss}, {Rogachevskii}, and
  {Sokoloff}}]{KleeorinMoss::2000}
{Kleeorin} N, {Moss} D, {Rogachevskii} I, et~al (2000) {Helicity balance and
  steady-state strength of the dynamo generated galactic magnetic field}. \aap
  361:L5--L8. \doi{10.48550/arXiv.astro-ph/0205266}

\bibitem[{{K{\"o}hler}(1973)}]{Koehler:1973}
{K{\"o}hler} H (1973) {The Solar Dynamo and Estimate of the Magnetic
  Diffusivity and the {\ensuremath{\alpha}}-effect}. \aap 25:467

\bibitem[{{Kotorashvili} et~al(2023){Kotorashvili}, {Blackman}, and
  {Owen}}]{Kotorashvili:etal:2023}
{Kotorashvili} K, {Blackman} EG, {Owen} JE (2023) {Why the observed spin
  evolution of older-than-solar like stars might not require a dynamo mode
  change}. arXiv e-prints arXiv:2301.04693.
  {\href{https://arxiv.org/abs/2301.04693}{{https://arxiv.org/abs/arXiv:2301.04693}}}
  {[astro-ph.SR]}

\bibitem[{{Kumar} et~al(2021){Kumar}, {Nagy}, {Lemerle}, {Karak}, and
  {Petrovay}}]{Kumar:etal:2021}
{Kumar} P, {Nagy} M, {Lemerle} A, et~al (2021) {The Polar Precursor Method for
  Solar Cycle Prediction: Comparison of Predictors and Their Temporal Range}.
  \apj 909(1):87. \doi{10.3847/1538-4357/abdbb4}

\bibitem[{{Kumar} et~al(2019){Kumar}, {Jouve}, and {Nandy}}]{Kumar:etal:2019}
{Kumar} R, {Jouve} L, {Nandy} D (2019) {A 3D kinematic Babcock Leighton solar
  dynamo model sustained by dynamic magnetic buoyancy and flux transport
  processes}. \aap 623:A54. \doi{10.1051/0004-6361/201834705}

\bibitem[{{Labonte} and {Howard}(1982)}]{LabonteHoward::1982}
{Labonte} BJ, {Howard} R (1982) {Torsional Waves on the Sun and the Activity
  Cycle}. \solphys 75(1-2):161--178. \doi{10.1007/BF00153469}

\bibitem[{{Larson} and {Schou}(2018)}]{LarsonSchou::2018}
{Larson} TP, {Schou} J (2018) {Global-Mode Analysis of Full-Disk Data from the
  Michelson Doppler Imager and the Helioseismic and Magnetic Imager}. \solphys
  293(2):29. \doi{10.1007/s11207-017-1201-5}

\bibitem[{{Legrand} and {Simon}(1981)}]{Legrand:Simon:1981}
{Legrand} JP, {Simon} PA (1981) {Ten Cycles of Solar and Geomagnetic Activity}.
  \solphys 70(1):173--195. \doi{https://doi.org/10.1007/BF00154399}

\bibitem[{{Leighton}(1964)}]{Leighton:1964}
{Leighton} RB (1964) {Transport of Magnetic Fields on the Sun.} \apj 140:1547.
  \doi{10.1086/148058}

\bibitem[{{Leighton}(1969)}]{Leighton:1969}
{Leighton} RB (1969) {A Magneto-Kinematic Model of the Solar Cycle}. \apj
  156:1. \doi{https://doi.org/10.1086/149943}

\bibitem[{{Lemerle} and {Charbonneau}(2017)}]{Lemerle:Charbonneau:2017}
{Lemerle} A, {Charbonneau} P (2017) {A Coupled 2 {\texttimes} 2D
  Babcock-Leighton Solar Dynamo Model. II. Reference Dynamo Solutions}. \apj
  834(2):133. \doi{10.3847/1538-4357/834/2/133}

\bibitem[{{Lemerle} et~al(2015){Lemerle}, {Charbonneau}, and
  {Carignan-Dugas}}]{Lemerle:etal:2015}
{Lemerle} A, {Charbonneau} P, {Carignan-Dugas} A (2015) {A Coupled 2
  {\texttimes} 2D Babcock-Leighton Solar Dynamo Model. I. Surface Magnetic Flux
  Evolution}. \apj 810(1):78. \doi{10.1088/0004-637X/810/1/78}

\bibitem[{{Liu} and {Scherrer}(2022)}]{LiuScherrer::2022}
{Liu} AL, {Scherrer} PH (2022) {Solar Toroidal Field Evolution Spanning Four
  Sunspot Cycles Seen by the Wilcox Solar Observatory, the Solar and
  Heliospheric Observatory/Michelson Doppler Imager, and the Solar Dynamics
  Observatory/Helioseismic and Magnetic Imager}. \apjl 927(1):L2.
  \doi{10.3847/2041-8213/ac52ae}

\bibitem[{{Longcope} and {Fisher}(1996)}]{LongcopeFisher::1996}
{Longcope} DW, {Fisher} GH (1996) {The Effects of Convection Zone Turbulence on
  the Tilt Angles of Magnetic Bipoles}. \apj 458:380. \doi{10.1086/176821}

\bibitem[{{Mackay} and {Yeates}(2012)}]{Mackay:Yeates:2012}
{Mackay} DH, {Yeates} AR (2012) {The Sun's Global Photospheric and Coronal
  Magnetic Fields: Observations and Models}. \lrsp 9(1):6.
  \doi{10.12942/lrsp-2012-6}

\bibitem[{{Mandal} et~al(2017){Mandal}, {Karak}, and
  {Banerjee}}]{Mandal:etal:2017}
{Mandal} S, {Karak} B, {Banerjee} D (2017) {Latitude Distribution of Sunspots:
  Analysis Using Sunspot Data and a Dynamo Model}. \apj 851(1):70.
  \doi{10.3847/1538-4357/aa97dc}

\bibitem[{{Martin} and {Harvey}(1979)}]{MartinHarvey::1979}
{Martin} SF, {Harvey} KH (1979) {Ephemeral Active Regions during Solar
  Minimum}. \solphys 64(1):93--108. \doi{10.1007/BF00151118}

\bibitem[{{Martin-Belda} and {Cameron}(2016)}]{Martin-Belda:Cameron:2016}
{Martin-Belda} D, {Cameron} RH (2016) {Surface flux transport simulations:
  Effect of inflows toward active regions and random velocities on the
  evolution of the Sun's large-scale magnetic field}. \aap 586:A73.
  \doi{10.1051/0004-6361/201527213}

\bibitem[{{Martin-Belda} and {Cameron}(2017)}]{Martin-BeldaCameron::2017}
{Martin-Belda} D, {Cameron} RH (2017) {Inflows towards active regions and the
  modulation of the solar cycle: A parameter study}. \aap 597:A21.
  \doi{10.1051/0004-6361/201629061}

\bibitem[{{McClintock} and {Norton}(2013)}]{McClintockNorton::2013}
{McClintock} BH, {Norton} AA (2013) {Recovering Joy's Law as a Function of
  Solar Cycle, Hemisphere, and Longitude}. \solphys 287(1-2):215--227.
  \doi{10.1007/s11207-013-0338-0}

\bibitem[{{McClintock} and {Norton}(2016)}]{McClintockNorton::2016}
{McClintock} BH, {Norton} AA (2016) {Tilt Angle and Footpoint Separation of
  Small and Large Bipolar Sunspot Regions Observed with HMI}. \apj 818(1):7.
  \doi{10.3847/0004-637X/818/1/7}

\bibitem[{{Metcalfe} et~al(2016){Metcalfe}, {Egeland}, and {van
  Saders}}]{MetcalfeEgeland::2016}
{Metcalfe} TS, {Egeland} R, {van Saders} J (2016) {Stellar Evidence That the
  Solar Dynamo May Be in Transition}. \apjl 826(1):L2.
  \doi{10.3847/2041-8205/826/1/L2}

\bibitem[{{Metcalfe} et~al(2022){Metcalfe}, {Finley}, {Kochukhov}, {See},
  {Ayres}, {Stassun}, {van Saders}, {Clark}, {Godoy-Rivera}, {Ilyin},
  {Pinsonneault}, {Strassmeier}, and {Petit}}]{Metcalfe:etal:2022}
{Metcalfe} TS, {Finley} AJ, {Kochukhov} O, et~al (2022) {The Origin of Weakened
  Magnetic Braking in Old Solar Analogs}. \apjl 933(1):L17.
  \doi{10.3847/2041-8213/ac794d}

\bibitem[{{Metcalfe} et~al(2023){Metcalfe}, {Strassmeier}, {Ilyin}, {van
  Saders}, {Ayres}, {Finley}, {Kochukhov}, {Petit}, {See}, {Stassun},
  {Jeffers}, {Marsden}, {Morin}, and {Vidotto}}]{Metcalfe:etal:2023}
{Metcalfe} TS, {Strassmeier} KG, {Ilyin} IV, et~al (2023) {Constraints on
  Magnetic Braking from the G8 Dwarf Stars 61 UMa and $\tau$ Cet}. arXiv
  e-prints arXiv:2304.09896. \doi{10.48550/arXiv.2304.09896}

\bibitem[{{Miesch} and {Dikpati}(2014)}]{Miesch:Dikpati:2014}
{Miesch} MS, {Dikpati} M (2014) {A Three-dimensional Babcock-Leighton Solar
  Dynamo Model}. \apjl 785(1):L8. \doi{10.1088/2041-8205/785/1/L8}

\bibitem[{{Moreno-Insertis} et~al(1992){Moreno-Insertis}, {Sch{\"u}ssler}, and
  {Ferriz-Mas}}]{Moreno-Insertis:etal:1992}
{Moreno-Insertis} F, {Sch{\"u}ssler} M, {Ferriz-Mas} A (1992) {Storage of
  magnetic flux tubes in a convective overshoot region}. \aap 264(2):686--700

\bibitem[{{Mu{\~n}oz-Jaramillo} et~al(2009){Mu{\~n}oz-Jaramillo}, {Nandy}, and
  {Martens}}]{Munoz-Jaramillo:etal:2009}
{Mu{\~n}oz-Jaramillo} A, {Nandy} D, {Martens} PCH (2009) {Helioseismic Data
  Inclusion in Solar Dynamo Models}. \apj 698(1):461--478.
  \doi{10.1088/0004-637X/698/1/461}

\bibitem[{{Mu{\~n}oz-Jaramillo} et~al(2010){Mu{\~n}oz-Jaramillo}, {Nandy},
  {Martens}, and {Yeates}}]{Munoz:Jaramillo:etal:2010}
{Mu{\~n}oz-Jaramillo} A, {Nandy} D, {Martens} PCH, et~al (2010) {A Double-ring
  Algorithm for Modeling Solar Active Regions: Unifying Kinematic Dynamo Models
  and Surface Flux-transport Simulations}. \apjl 720(1):L20--L25.
  \doi{10.1088/2041-8205/720/1/L20}

\bibitem[{{Mu{\~n}oz-Jaramillo} et~al(2013){Mu{\~n}oz-Jaramillo},
  {Dasi-Espuig}, {Balmaceda}, and {DeLuca}}]{Munoz-Jaramillo:etal:2013}
{Mu{\~n}oz-Jaramillo} A, {Dasi-Espuig} M, {Balmaceda} LA, et~al (2013) {Solar
  Cycle Propagation, Memory, and Prediction: Insights from a Century of
  Magnetic Proxies}. \apjl 767(2):L25. \doi{10.1088/2041-8205/767/2/L25}

\bibitem[{{Nagy} et~al(2017){Nagy}, {Lemerle}, {Labonville}, {Petrovay}, and
  {Charbonneau}}]{NagyLemerle::2017}
{Nagy} M, {Lemerle} A, {Labonville} F, et~al (2017) {The Effect of ``Rogue''
  Active Regions on the Solar Cycle}. \solphys 292(11):167.
  \doi{10.1007/s11207-017-1194-0}

\bibitem[{{Nagy} et~al(2020){Nagy}, {Lemerle}, and
  {Charbonneau}}]{NagyLemerle::2020}
{Nagy} M, {Lemerle} A, {Charbonneau} P (2020) {Impact of nonlinear surface
  inflows into activity belts on the solar dynamo}. Journal of Space Weather
  and Space Climate 10:62. \doi{10.1051/swsc/2020064}

\bibitem[{{Nandy} and {Choudhuri}(2001)}]{Nandy:Choudhuri:2001}
{Nandy} D, {Choudhuri} AR (2001) {Toward a Mean Field Formulation of the
  Babcock-Leighton Type Solar Dynamo. I. {\ensuremath{\alpha}}-Coefficient
  versus Durney's Double-Ring Approach}. \apj 551(1):576--585.
  \doi{10.1086/320057}

\bibitem[{{Nandy} and {Choudhuri}(2002)}]{Nandy:Choudhuri:2002}
{Nandy} D, {Choudhuri} AR (2002) {Explaining the Latitudinal Distribution of
  Sunspots with Deep Meridional Flow}. Science 296(5573):1671--1673.
  \doi{10.1126/science.1070955}

\bibitem[{{Nelson} et~al(2014){Nelson}, {Brown}, {Sacha Brun}, {Miesch}, and
  {Toomre}}]{Nelson:etal:2014}
{Nelson} NJ, {Brown} BP, {Sacha Brun} A, et~al (2014) {Buoyant Magnetic Loops
  Generated by Global Convective Dynamo Action}. \solphys 289(2):441--458.
  \doi{10.1007/s11207-012-0221-4}

\bibitem[{{Parker}(1955{\natexlab{a}})}]{Parker:1955b}
{Parker} EN (1955{\natexlab{a}}) {Hydromagnetic Dynamo Models.} \apj 122:293.
  \doi{10.1086/146087}

\bibitem[{{Parker}(1955{\natexlab{b}})}]{Parker:1955a}
{Parker} EN (1955{\natexlab{b}}) {The Formation of Sunspots from the Solar
  Toroidal Field.} \apj 121:491. \doi{10.1086/146010}

\bibitem[{{Parker}(1975)}]{Parker:1975}
{Parker} EN (1975) {The generation of magnetic fields in astrophysical bodies.
  X. Magnetic buoyancy and the solar dynamo.} \apj 198:205--209.
  \doi{10.1086/153593}

\bibitem[{{Parker}(1984)}]{Parker::1984}
{Parker} EN (1984) {Magnetic buoyancy and the escape of magnetic fields from
  stars}. \apj 281:839--845. \doi{10.1086/162163}

\bibitem[{{Parker}(1993)}]{Parker:1993}
{Parker} EN (1993) {A Solar Dynamo Surface Wave at the Interface between
  Convection and Nonuniform Rotation}. \apj 408:707. \doi{10.1086/172631}

\bibitem[{{Parker}(2009)}]{Parker:2009}
{Parker} EN (2009) {Solar Magnetism: The State of Our Knowledge and Ignorance}.
  \ssr 144(1-4):15--24. \doi{10.1007/s11214-008-9445-x}

\bibitem[{{Petrovay}(2020)}]{Petrovay:2020}
{Petrovay} K (2020) {Solar cycle prediction}. \lrsp 17(1):2.
  \doi{10.1007/s41116-020-0022-z}

\bibitem[{{Petrovay} et~al(2020){Petrovay}, {Nagy}, and
  {Yeates}}]{PetrovayNagy::2020}
{Petrovay} K, {Nagy} M, {Yeates} AR (2020) {Towards an algebraic method of
  solar cycle prediction. I. Calculating the ultimate dipole contributions of
  individual active regions}. Journal of Space Weather and Space Climate 10:50.
  \doi{10.1051/swsc/2020050}

\bibitem[{{Pipin}(2022)}]{Pipin:2022}
{Pipin} VV (2022) {On the effect of surface bipolar magnetic regions on the
  convection zone dynamo}. \mnras 514(1):1522--1534.
  \doi{10.1093/mnras/stac1434}

\bibitem[{{Reiners} et~al(2022){Reiners}, {Shulyak}, {K{\"a}pyl{\"a}}, {Ribas},
  {Nagel}, {Zechmeister}, {Caballero}, {Shan}, {Fuhrmeister}, {Quirrenbach},
  {Amado}, {Montes}, {Jeffers}, {Azzaro}, {B{\'e}jar}, {Chaturvedi}, {Henning},
  {K{\"u}rster}, and {Pall{\'e}}}]{Reiners:etal:2022}
{Reiners} A, {Shulyak} D, {K{\"a}pyl{\"a}} PJ, et~al (2022) {Magnetism,
  rotation, and nonthermal emission in cool stars. Average magnetic field
  measurements in 292 M dwarfs}. \aap 662:A41.
  \doi{10.1051/0004-6361/202243251}

\bibitem[{{Rempel}(2006)}]{Rempel:2006}
{Rempel} M (2006) {Flux-Transport Dynamos with Lorentz Force Feedback on
  Differential Rotation and Meridional Flow: Saturation Mechanism and Torsional
  Oscillations}. \apj 647(1):662--675. \doi{10.1086/505170}

\bibitem[{{Route}(2016)}]{Route:2016}
{Route} M (2016) {The Discovery of Solar-like Activity Cycles Beyond the End of
  the Main Sequence?} \apjl 830(2):L27. \doi{10.3847/2041-8205/830/2/L27}

\bibitem[{{R{\"u}diger} and {Brandenburg}(1995)}]{Ruediger:Brandenburg:1995}
{R{\"u}diger} G, {Brandenburg} A (1995) {A solar dynamo in the overshoot layer:
  cycle period and butterfly diagram.} \aap 296:557

\bibitem[{{Schatten} et~al(1978){Schatten}, {Scherrer}, {Svalgaard}, and
  {Wilcox}}]{Schatten:etal:1978}
{Schatten} KH, {Scherrer} PH, {Svalgaard} L, et~al (1978) {Using Dynamo Theory
  to predict the sunspot number during Solar Cycle 21}. \grl 5(5):411--414.
  \doi{10.1029/GL005i005p00411}

\bibitem[{{Schunker} et~al(2016){Schunker}, {Braun}, {Birch}, {Burston}, and
  {Gizon}}]{SchunkerBraun::2016}
{Schunker} H, {Braun} DC, {Birch} AC, et~al (2016) {SDO/HMI survey of emerging
  active regions for helioseismology}. \aap 595:A107.
  \doi{10.1051/0004-6361/201628388}

\bibitem[{{Schunker} et~al(2019){Schunker}, {Birch}, {Cameron}, {Braun},
  {Gizon}, and {Burston}}]{Schunker:etal:2019}
{Schunker} H, {Birch} AC, {Cameron} RH, et~al (2019) {Average motion of
  emerging solar active region polarities. I. Two phases of emergence}. \aap
  625:A53. \doi{10.1051/0004-6361/201834627}

\bibitem[{{Schunker} et~al(2020){Schunker}, {Baumgartner}, {Birch}, {Cameron},
  {Braun}, and {Gizon}}]{Schunker:etal:2020}
{Schunker} H, {Baumgartner} C, {Birch} AC, et~al (2020) {Average motion of
  emerging solar active region polarities. II. Joy's law}. \aap 640:A116.
  \doi{10.1051/0004-6361/201937322}

\bibitem[{{Sch{\"u}ssler} and {Rempel}(2005)}]{Schuessler:Rempel:2005}
{Sch{\"u}ssler} M, {Rempel} M (2005) {The dynamical disconnection of sunspots
  from their magnetic roots}. \aap 441(1):337--346.
  \doi{10.1051/0004-6361:20052962}

\bibitem[{{Sch{\"u}ssler} et~al(1994){Sch{\"u}ssler}, {Caligari}, {Ferriz-Mas},
  and {Moreno-Insertis}}]{Schuessler:etal:1994}
{Sch{\"u}ssler} M, {Caligari} P, {Ferriz-Mas} A, et~al (1994) {Instability and
  eruption of magnetic flux tubes in the solar convection zone.} \aap
  281:L69--L72

\bibitem[{{Solanki} et~al(2004){Solanki}, {Usoskin}, {Kromer}, {Sch{\"u}ssler},
  and {Beer}}]{SolankiUsoskin::2004}
{Solanki} SK, {Usoskin} IG, {Kromer} B, et~al (2004) {Unusual activity of the
  Sun during recent decades compared to the previous 11,000 years}. \nat
  431(7012):1084--1087. \doi{10.1038/nature02995}

\bibitem[{{Solanki} et~al(2008){Solanki}, {Wenzler}, and
  {Schmitt}}]{Solanki:etal:2008}
{Solanki} SK, {Wenzler} T, {Schmitt} D (2008) {Moments of the latitudinal
  dependence of the sunspot cycle: a new diagnostic of dynamo models}. \aap
  483(2):623--632. \doi{10.1051/0004-6361:20054282}

\bibitem[{{Spruit}(1997)}]{Spruit:1997}
{Spruit} HC (1997) {Convection in stellar envelopes: a changing paradigm.}
  Memorie della Societa Astronomica Italiana 68:397--413.
  \doi{10.48550/arXiv.astro-ph/9605020}

\bibitem[{{Spruit}(2011)}]{Spruit:2011}
{Spruit} HC (2011) {Theories of the Solar Cycle: A Critical View}. In:
  {Miralles} MP, {S{\'a}nchez Almeida} J (eds) The Sun, the Solar Wind, and the
  Heliosphere, vol~4. Springer, Dordrecht, p~39

\bibitem[{{Spruit}(2012)}]{Spruit:2012}
{Spruit} HC (2012) {Theories of the Solar Cycle and Its Effect on Climate}.
  Progress of Theoretical Physics Supplement 195:185--200.
  \doi{10.1143/PTPS.195.185}

\bibitem[{{Spruit} and {van Ballegooijen}(1982)}]{Spruit:Ballegooijen:1982}
{Spruit} HC, {van Ballegooijen} AA (1982) {Stability of toroidal flux tubes in
  stars}. \aap 106(1):58--66

\bibitem[{{Steenbeck} and {Krause}(1966)}]{Steenbeck:Krause:1966}
{Steenbeck} M, {Krause} F (1966) {Erkl{\"a}rung stellarer und planetarer
  Magnetfelder durch einen turbulenzbedingten Dynamomechanismus}. Zs
  Naturforsch A 21:1285. \doi{10.1515/zna-1966-0813}

\bibitem[{{Steenbeck} and {Krause}(1969)}]{SteenbeckKrause::1969}
{Steenbeck} M, {Krause} F (1969) {On the Dynamo Theory of Stellar and Planetary
  Magnetic Fields. I. AC Dynamos of Solar Type}. \an 291:49--84.
  \doi{10.1002/asna.19692910201}

\bibitem[{{Steenbeck} et~al(1966){Steenbeck}, {Krause}, and
  {R{\"a}dler}}]{Steenbeck:etal:1966}
{Steenbeck} M, {Krause} F, {R{\"a}dler} KH (1966) {Berechnung der mittleren
  Lorentz-Feldst{\"a}rke f{\"u}r ein elektrisch leitendes Medium in
  turbulenter, durch Coriolis-Kr{\"a}fte beeinflu{\ss}ter Bewegung}.
  Zeitschrift Naturforschung Teil A 21:369. \doi{10.1515/zna-1966-0401}

\bibitem[{{Stix}(1972)}]{Stix::1972}
{Stix} M (1972) {Non-Linear Dynamo Waves}. \aap 20:9

\bibitem[{{Talafha} et~al(2022){Talafha}, {Nagy}, {Lemerle}, and
  {Petrovay}}]{Talafha:etal:2022}
{Talafha} M, {Nagy} M, {Lemerle} A, et~al (2022) {Role of observable
  nonlinearities in solar cycle modulation}. \aap 660:A92.
  \doi{10.1051/0004-6361/202142572}

\bibitem[{{Tobias}(1996)}]{Tobias:1996}
{Tobias} SM (1996) {Diffusivity Quenching as a Mechanism for Parker's Surface
  Dynamo}. \apj 467:870. \doi{10.1086/177661}

\bibitem[{{Tobias} et~al(1995){Tobias}, {Weiss}, and
  {Kirk}}]{TobiasWeiss::1995}
{Tobias} SM, {Weiss} NO, {Kirk} V (1995) {Chaotically modulated stellar
  dynamos}. \mnras 273(4):1150--1166. \doi{10.1093/mnras/273.4.1150}

\bibitem[{{Tripathi} et~al(2021){Tripathi}, {Nandy}, and
  {Banerjee}}]{Tripathi:etal:2021}
{Tripathi} B, {Nandy} D, {Banerjee} S (2021) {Stellar mid-life crisis:
  subcritical magnetic dynamos of solar-like stars and the breakdown of
  gyrochronology}. \mnras 506(1):L50--L54. \doi{10.1093/mnrasl/slab035}

\bibitem[{{Upton} and {Hathaway}(2014)}]{Upton:Hathaway:2014}
{Upton} L, {Hathaway} DH (2014) {Predicting the Sun's Polar Magnetic Fields
  with a Surface Flux Transport Model}. \apj 780(1):5.
  \doi{10.1088/0004-637X/780/1/5}

\bibitem[{{Usoskin}(2017)}]{Usoskin:2017}
{Usoskin} IG (2017) {A history of solar activity over millennia}. \lrsp
  14(1):3. \doi{10.1007/s41116-017-0006-9}

\bibitem[{{Usoskin} et~al(2016){Usoskin}, {Gallet}, {Lopes}, {Kovaltsov}, and
  {Hulot}}]{UsoskinGallet::2016}
{Usoskin} IG, {Gallet} Y, {Lopes} F, et~al (2016) {Solar activity during the
  Holocene: the Hallstatt cycle and its consequence for grand minima and
  maxima}. \aap 587:A150. \doi{10.1051/0004-6361/201527295}

\bibitem[{{van Saders} et~al(2016){van Saders}, {Ceillier}, {Metcalfe}, {Silva
  Aguirre}, {Pinsonneault}, {Garc{\'\i}a}, {Mathur}, and
  {Davies}}]{vanSadersCeillier::2016}
{van Saders} JL, {Ceillier} T, {Metcalfe} TS, et~al (2016) {Weakened magnetic
  braking as the origin of anomalously rapid rotation in old field stars}. \nat
  529(7585):181. \doi{10.1038/nature16168}

\bibitem[{{Waldmeier}(1939)}]{Waldmeier:1939}
{Waldmeier} M (1939) {Die Zonenwanderung der Sonnenflecken}. \meth 14:470--481

\bibitem[{{Waldmeier}(1955)}]{Waldmeier:1955}
{Waldmeier} M (1955) {Ergebnisse und Probleme der Sonnenforschung.}

\bibitem[{{Wallenhorst} and {Topka}(1982)}]{Wallenhorst:Topka:1982}
{Wallenhorst} SG, {Topka} KP (1982) {On the Disappearance of a Small Sunspot
  Group}. \solphys 81(1):33--46. \doi{10.1007/BF00151977}

\bibitem[{{Wang} and {Sheeley}(1991)}]{Wang:Sheeley:1991}
{Wang} YM, {Sheeley} JN.~R. (1991) {Magnetic Flux Transport and the Sun's
  Dipole Moment: New Twists to the Babcock-Leighton Model}. \apj 375:761.
  \doi{10.1086/170240}

\bibitem[{{Wang} and {Sheeley}(2009)}]{Wang:Sheeley:2009}
{Wang} YM, {Sheeley} NR (2009) {Understanding the Geomagnetic Precursor of the
  Solar Cycle}. \apjl 694(1):L11--L15. \doi{10.1088/0004-637X/694/1/L11}

\bibitem[{{Wang} et~al(1989{\natexlab{a}}){Wang}, {Nash}, and
  {Sheeley}}]{Wang:etal:1989b}
{Wang} YM, {Nash} AG, {Sheeley} JN.~R. (1989{\natexlab{a}}) {Evolution of the
  Sun's Polar Fields during Sunspot Cycle 21: Poleward Surges and Long-Term
  Behavior}. \apj 347:529. \doi{10.1086/168143}

\bibitem[{{Wang} et~al(1989{\natexlab{b}}){Wang}, {Nash}, and
  {Sheeley}}]{Wang:etal:1989a}
{Wang} YM, {Nash} AG, {Sheeley} JN.~R. (1989{\natexlab{b}}) {Magnetic Flux
  Transport on the Sun}. Science 245(4919):712--718.
  \doi{10.1126/science.245.4919.712}

\bibitem[{{Wang} et~al(1991){Wang}, {Sheeley}, and {Nash}}]{Wang:etal:1991}
{Wang} YM, {Sheeley} JN.~R., {Nash} AG (1991) {A New Solar Cycle Model
  Including Meridional Circulation}. \apj 383:431. \doi{10.1086/170800}

\bibitem[{{Wang} et~al(2002){Wang}, {Sheeley}, and {Lean}}]{Wang:etal:2002}
{Wang} YM, {Sheeley} JN.~R., {Lean} J (2002) {Meridional Flow and the Solar
  Cycle Variation of the Sun's Open Magnetic Flux}. \apj 580(2):1188--1196.
  \doi{10.1086/343845}

\bibitem[{{Whitbread} et~al(2017){Whitbread}, {Yeates}, {Mu{\~n}oz-Jaramillo},
  and {Petrie}}]{Whitbread:etal:2017}
{Whitbread} T, {Yeates} AR, {Mu{\~n}oz-Jaramillo} A, et~al (2017) {Parameter
  optimization for surface flux transport models}. \aap 607:A76.
  \doi{10.1051/0004-6361/201730689}

\bibitem[{{Whitbread} et~al(2018){Whitbread}, {Yeates}, and
  {Mu{\~n}oz-Jaramillo}}]{WhitbreadYeates::2018}
{Whitbread} T, {Yeates} AR, {Mu{\~n}oz-Jaramillo} A (2018) {How Many Active
  Regions Are Necessary to Predict the Solar Dipole Moment?} \apj 863(2):116.
  \doi{10.3847/1538-4357/aad17e}

\bibitem[{{Whitbread} et~al(2019){Whitbread}, {Yeates}, and
  {Mu{\~n}oz-Jaramillo}}]{Whitbread:etal:2019}
{Whitbread} T, {Yeates} AR, {Mu{\~n}oz-Jaramillo} A (2019) {The need for active
  region disconnection in 3D kinematic dynamo simulations}. \aap 627:A168.
  \doi{10.1051/0004-6361/201935986}

\bibitem[{{Wilson} et~al(1988){Wilson}, {Altrocki}, {Harvey}, {Martin}, and
  {Snodgrass}}]{WilsonAltrocki::1988}
{Wilson} PR, {Altrocki} RC, {Harvey} KL, et~al (1988) {The extended solar
  activity cycle}. \nat 333(6175):748--750. \doi{10.1038/333748a0}

\bibitem[{{Wright} and {Drake}(2016)}]{Wright:Drake:2016}
{Wright} NJ, {Drake} JJ (2016) {Solar-type dynamo behaviour in fully convective
  stars without a tachocline}. \nat 535(7613):526--528.
  \doi{10.1038/nature18638}

\bibitem[{{Yeates} and
  {Mu{\~n}oz-Jaramillo}(2013)}]{Yeates:Munoz-Jaramillo:2013}
{Yeates} AR, {Mu{\~n}oz-Jaramillo} A (2013) {Kinematic active region formation
  in a three-dimensional solar dynamo model}. \mnras 436(4):3366--3379.
  \doi{10.1093/mnras/stt1818}

\bibitem[{{Yeates} et~al(2008){Yeates}, {Nandy}, and
  {Mackay}}]{Yeates:etal:2008}
{Yeates} AR, {Nandy} D, {Mackay} DH (2008) {Exploring the Physical Basis of
  Solar Cycle Predictions: Flux Transport Dynamics and Persistence of Memory in
  Advection- versus Diffusion-dominated Solar Convection Zones}. \apj
  673(1):544--556. \doi{10.1086/524352}

\bibitem[{{Yeates} et~al(2023){Yeates}, {Cheung}, {Jiang}, {Petrovay}, and
  {Wang}}]{Yeates:etal:2023}
{Yeates} AR, {Cheung} MCM, {Jiang} J, et~al (2023) {Surface Flux Transport}.
  \ssr p submitted. \doi{10.48550/arXiv.2303.01209}

\bibitem[{{Yoshimura}(1975)}]{Yoshimura:1975}
{Yoshimura} H (1975) {Solar-cycle dynamo wave propagation.} \apj 201:740--748.
  \doi{10.1086/153940}

\bibitem[{{Yule}(1927)}]{Yule:1927}
{Yule} GU (1927) {On a Method of Investigating Periodicities in Disturbed
  Series, with special reference to Wolfer's Sunspot Numbers}. Phil Trans Roy
  Soc London A 226:267--298

\bibitem[{{Zhang} and {Jiang}(2022)}]{Zhang:Jiang:2022}
{Zhang} Z, {Jiang} J (2022) {A Babcock-Leighton-type Solar Dynamo Operating in
  the Bulk of the Convection Zone}. \apj 930(1):30.
  \doi{10.3847/1538-4357/ac6177}

\end{thebibliography}

\end{document}